\documentclass[prd,twocolumn,reprint,superscriptaddress,groupedaddress]{revtex4}  
\usepackage{graphicx}  
\usepackage{dcolumn}   
\usepackage{bm}        
\usepackage{amssymb, amsmath}
\usepackage{subcaption}
\usepackage{color,soul}
\usepackage{float}
\usepackage{upgreek}
\usepackage{relsize}
\def\textsubscript#1{\ensuremath{_{\mbox{\textscale{.6}{#1}}}}}

\newcommand{\param}{\ensuremath{\vec{\theta}} }

\newcommand\abs[1]{\left|#1\right|}
\newcommand{\hd}{\hphantom{0}}
\begin{document}

\newcommand\tma{TM\textsubscript{1}\,}
\newcommand\tmb{TM\textsubscript{2}\,}

\widetext

\title{Calibrating the System Dynamics of LISA Pathfinder}
\def\addressa{European Space Astronomy Centre, European Space Agency, Villanueva de la
Ca\~{n}ada, 28692 Madrid, Spain}
\def\addressb{Albert-Einstein-Institut, Max-Planck-Institut f\"ur Gravitationsphysik und Leibniz Universit\"at Hannover,
Callinstra{\ss}e 38, 30167 Hannover, Germany}
\def\addressc{APC, Univ Paris Diderot, CNRS/IN2P3, CEA/lrfu, Obs de Paris, Sorbonne Paris Cit\'e 75013, France}
\def\addressd{High Energy Physics Group, Physics Department, Imperial College London, Blackett Laboratory, Prince Consort Road, London, SW7 2BW, UK }
\def\addresse{Dipartimento di Fisica, Universit\`a di Roma ``Tor Vergata'',  and INFN, sezione Roma Tor Vergata, I-00133 Roma, Italy}
\def\addressf{Department of Industrial Engineering, University of Trento, via Sommarive 9, 38123 Trento, 
and Trento Institute for Fundamental Physics and Application / INFN}
\def\addressh{European Space Technology Centre, European Space Agency, 
Keplerlaan 1, 2200 AG Noordwijk, The Netherlands}
\def\addressi{Dipartimento di Fisica, Universit\`a di Trento and Trento Institute for 
Fundamental Physics and Application / INFN, 38123 Povo, Trento, Italy}
\def\addressk{Istituto di Fotonica e Nanotecnologie, CNR-Fondazione Bruno Kessler, I-38123 Povo, Trento, Italy}
\def\addressj{The School of Physics and Astronomy, University of
Birmingham, Birmingham B15 2TT, UK}
\def\addressl{Institut f\"ur Geophysik, ETH Z\"urich, Sonneggstrasse 5, CH-8092, Z\"urich, Switzerland}
\def\addressm{The UK Astronomy Technology Centre, Royal Observatory, Edinburgh, Blackford Hill, Edinburgh, EH9 3HJ, UK}
\def\addressn{Institut de Ci\`encies de l'Espai (CSIC-IEEC), Campus UAB, Carrer de Can Magrans s/n, 08193 Cerdanyola del Vall\`es, Spain}
\def\addresso{DISPEA, Universit\`a di Urbino ``Carlo Bo'', Via S. Chiara, 27 61029 Urbino/INFN, Italy}
\def\addressp{European Space Operations Centre, European Space Agency, 64293 Darmstadt, Germany }
\def\addressq{Physik Institut, 
Universit\"at Z\"urich, Winterthurerstrasse 190, CH-8057 Z\"urich, Switzerland}
\def\addressr{SUPA, Institute for Gravitational Research, School of Physics and Astronomy, University of Glasgow, Glasgow, G12 8QQ, UK}
\def\addresss{Department d'Enginyeria Electr\`onica, Universitat Polit\`ecnica de Catalunya,  08034 Barcelona, Spain}
\def\addresst{Institut d'Estudis Espacials de Catalunya (IEEC), C/ Gran Capit\`a 2-4, 08034 Barcelona, Spain}
\def\addressu{Gravitational Astrophysics Lab, NASA Goddard Space Flight Center, 8800 Greenbelt Road, Greenbelt, MD 20771 USA}
\def\addressbb{Department of Physics, University of Florida, 2001 Museum Rd, Gainesville, FL 32603, USA}
\def\addresscc{Istituto di Fotonica e Nanotecnologie, CNR-Fondazione Bruno Kessler, I-38123 Povo, Trento, Italy}

\author{M~Armano}\affiliation{\addressa}
\author{H~Audley}\affiliation{\addressb}
\author{J~Baird}\affiliation{\addressd}
\author{P~Binetruy}\affiliation{\addressc}
\author{M~Born}\affiliation{\addressb}
\author{D~Bortoluzzi}\affiliation{\addressf}
\author{E~Castelli}\affiliation{\addressi}
\author{A~Cavalleri}\affiliation{\addresscc}
\author{A~Cesarini}\affiliation{\addresso}
\author{A\,M~Cruise}\affiliation{\addressj}
\author{K~Danzmann}\affiliation{\addressb}
\author{M~de Deus Silva}\affiliation{\addressa}
\author{I~Diepholz}\affiliation{\addressb}
\author{G~Dixon}\affiliation{\addressj}
\author{R~Dolesi}\affiliation{\addressi}
\author{L~Ferraioli}\affiliation{\addressl}
\author{V~Ferroni}\affiliation{\addressi}
\author{E\,D~Fitzsimons}\affiliation{\addressm}
\author{M~Freschi}\affiliation{\addressa}
\author{L~Gesa}\affiliation{\addressn}
\author{F~Gibert}\affiliation{\addressi}
\author{D~Giardini}\affiliation{\addressl}
\author{R~Giusteri}\affiliation{\addressi}
\author{C~Grimani}\affiliation{\addresso}
\author{J~Grzymisch}\affiliation{\addressh}
\author{I~Harrison}\affiliation{\addressp}
\author{G~Heinzel}\affiliation{\addressb}
\author{M~Hewitson}\affiliation{\addressb}
\author{D~Hollington}\affiliation{\addressd}
\author{D~Hoyland}\affiliation{\addressj}
\author{M~Hueller}\affiliation{\addressi}
\author{H~Inchausp\'e}\affiliation{\addressc}
\author{O~Jennrich}\affiliation{\addressh}
\author{P~Jetzer}\affiliation{\addressq}
\author{N~Karnesis}\email{karnesis@aei.mpg.de}\affiliation{\addressb}\affiliation{\addressc}
\author{B~Kaune}\affiliation{\addressb}
\author{N~Korsakova}\affiliation{\addressr}
\author{C\,J~Killow}\affiliation{\addressr}
\author{J\,A~Lobo}\affiliation{\addressn}
\author{I~Lloro}\affiliation{\addressn}
\author{L~Liu}\affiliation{\addressi}
\author{J\,P~L\'opez-Zaragoza}\affiliation{\addressn}
\author{R~Maarschalkerweerd}\affiliation{\addressp}
\author{D~Mance}\affiliation{\addressl}
\author{N~Meshksar}\affiliation{\addressl}
\author{V~Mart\'{i}n}\affiliation{\addressn}
\author{L~Martin-Polo}\affiliation{\addressa}
\author{J~Martino}\affiliation{\addressc}
\author{F~Martin-Porqueras}\affiliation{\addressa}
\author{I~Mateos}\affiliation{\addressn}
\author{P\,W~McNamara}\affiliation{\addressh}
\author{J~Mendes}\affiliation{\addressp}
\author{L~Mendes}\affiliation{\addressa}
\author{M~Nofrarias}\affiliation{\addressn}
\author{S~Paczkowski}\affiliation{\addressb}
\author{M~Perreur-Lloyd}\affiliation{\addressr}
\author{A~Petiteau}\affiliation{\addressc}
\author{P~Pivato}\affiliation{\addressi}
\author{E~Plagnol}\affiliation{\addressc}
\author{J~Ramos-Castro}\affiliation{\addresss}
\author{J~Reiche}\affiliation{\addressb}
\author{D\,I~Robertson}\affiliation{\addressr}
\author{F~Rivas}\affiliation{\addressn}
\author{G~Russano}\affiliation{\addressi}
\author{J~Slutsky}\affiliation{\addressu}
\author{C\,F~Sopuerta}\affiliation{\addressn}
\author{T~Sumner}\affiliation{\addressd}
\author{D~Texier}\affiliation{\addressa}
\author{J\,I~Thorpe}\affiliation{\addressu}
\author{D~Vetrugno}\email{daniele.vetrugno@unitn.it}\affiliation{\addressi}
\author{S~Vitale}\affiliation{\addressi}
\author{G~Wanner}\affiliation{\addressb}
\author{H~Ward}\affiliation{\addressr}
\author{P~Wass}\affiliation{\addressd}\affiliation{\addressbb}
\author{W\,J~Weber}\affiliation{\addressi}
\author{L~Wissel}\affiliation{\addressb}
\author{A~Wittchen}\affiliation{\addressb}
\author{P~Zweifel}\affiliation{\addressl}


\begin{abstract}

LISA Pathfinder (LPF) was a European Space Agency mission
with the aim to test key technologies for future
space-borne gravitational-wave observatories like LISA. The main scientific goal of LPF was to demonstrate measurements of 
differential acceleration between free-falling test masses at the sub-femto-$g$ level, and to understand the  residual acceleration 
in terms of a physical model of stray forces, and displacement readout noise. A key step 
towards reaching the LPF goals was the correct calibration of the dynamics of LPF, which was a three-body
system composed by two test-masses enclosed in a single spacecraft, and subject to control laws for system
stability. In this work, we report on the calibration procedures adopted to calculate the residual differential
stray force per unit mass acting on the two test-masses in their nominal positions. 
The physical parameters of the adopted dynamical model are
presented, together with their role on LPF performance. 
The analysis and results of these experiments show that the dynamics of the system was accurately modeled and the dynamical parameters were stationary throughout the mission.
Finally, the impact and importance of calibrating system dynamics for future space-based gravitational wave observatories is discussed.

\end{abstract}

\maketitle

\section{Introduction
\label{section:introduction}}

The LISA Pathfinder satellite~\cite{0264-9381-25-11-114034, lpf1} (LPF) was a technology demonstrator for future
space-borne Gravitational Wave (GW) observatories, such as LISA~\cite{2017arXiv170200786A}. It aimed to
measure  the relative displacement $\Delta x(t)$ between two freely falling test masses using heterodyne laser interferometry, and
to demonstrate that the fluctuations of the differential stray force per unit mass,
$\Delta g(t)$, acting on the two TMs in their nominal positions
were below $30\,\mathrm{fm\,s^{-2}}/\sqrt{\text{Hz}}$ in the millihertz frequency band. Such a level of residual
acceleration would allow LISA to detect and distinguish the GW signals originating from 
sources in the $0.1$~mHz to $100$~mHz band~\cite{2017arXiv170200786A}. 
The main objective of the LPF mission was not to detect GW but rather to test key
technologies for LISA, and to measure, model and eventually subtract the main contributions to $\Delta g(t)$.

The LPF satellite was launched on the 3rd of December, 2015 to the L1 Lagrange point, with its
science operations starting on March 1st, 2016. During its first four month of science operations, 
called nominal phase, it showed an unprecedented performance of differential acceleration noise, 
reaching Amplitude Spectral Density (ASD) levels of $S_{\Delta g}^{1/2}\simeq 5.57\pm0.04 \,\mathrm{fm\,s^{-2}}/\sqrt{\text{Hz}}$ 
between 1 and 10~mHz~\cite{lpf_prl}, well below the primary goal of the
mission. Nominal operations ended on June 25th, 2016, when the NASA experiments on the spacecraft 
commenced operations. In this second phase, guided by NASA scientists, a set of colloidal thrusters, 
instead of the cold gas thrusters used during the nominal phase, were used to control the spacecraft 
and several experiments were performed in order to characterize the behaviour of those new 
thrusters. On the 7th of December 2016 NASA operations finished, and an LPF mission extension 
phase started, demonstrating an even better differential acceleration noise performance reaching $S_{\Delta g}^{1/2}\simeq 1.74\pm0.05 \,\mathrm{fm\,s^{-2}}/\sqrt{\text{Hz}}$ above $2~\text{mHz}$ and $S_{\Delta g}^{1/2}\simeq (6\pm1)\times10 \,\mathrm{fm\,s^{-2}}/\sqrt{\text{Hz}}$ at $20~\mu\text{Hz}$~\cite{lpf_prl2}. The 
extension phase ran until the 18th of July 2017 when the satellite was finally passivated.

Accurate identification and calibration of the system dynamics of LPF were vital to the understanding and optimization of the 
measurement of $\Delta g(t)$. Indeed, the LPF satellite requires control loops to keep operating in a stable configuration and for this on-purpose forces are constantly applied on one TM and the spacecraft. The 
consequent forces per unit mass perturb the geodesic motion of the TMs. Furthermore, the local gradient fields surrounding the 
TMs, generate forces per unit mass by means of their coupling to the TMs displacement, which again perturb the geodesic motion of the TMs. All those effects can be estimated and subtracted. For 
this purpose, during the mission, a series of experiments to estimate the dynamical free 
parameters of LPF were designed and performed on board. They
consisted of injections of high signal-to-noise-ratio calibration modulations of the dynamical control set-points, in order to excite the 
relative motion of the test masses and spacecraft. While their design aimed to
characterize the dynamical environment of the LPF, their secondary purpose was to investigate the stability and
stationarity of the controllers and actuators. 

A further correction of the $\Delta g$ data was required for effects of the non-inertial platform at very low frequencies. Measurements of $\Delta x(t)$ are performed with an optical bench attached to the spacecraft frame, whose rotational motion introduces fictitious forces, both due to misalignments and to the centrifugal force, most relevantly at frequencies around 0.1 mHz. Correction for these effects substantially improved the noise performance of the instrument.

In this work, we report on the procedure and methodology that were followed in order to derive the $\Delta g(t)$
quantity, and we discuss the impact of this measurement to the case of a grand-scale observatory
such as LISA. Section~\ref{section:dynamics} describes the dynamics of the system and introduces all the possible dynamical sources of
force noise that contribute to the overall $\Delta g(t)$ noise budget. In section~\ref{section:sysid} the 
system identification experiments performed on-board the satellite are described, and the data analysis methodology is 
explained. We present two approaches; First, we overview the computationally cheap method we utilized during operations, 
in order to get a first reliable estimation of the dynamics of the system. Secondly, we 
report on the technique employed to perform a global fit on the complete data set of the system identification experiments performed
during the mission. In this case, we define a model of the forces gradient inside the satellite, which depends on the various 
configurations of the instrument. This approach aims to estimate the background components of that gradient which can not be modelled and parametrized. The experiments reported here are those performed along the sensitive $x$-axis joining the two test
masses, the only one relevant for the requirements of $\Delta g(t)$. Section~\ref{section:inertial} introduces and explains 
the complementary and independent low frequency calibration of $\Delta g(t)$, related to the appearance of inertial forces in  the rotating frame of LPF.
Section~\ref{section:results} presents the main results of all performed calibrations, and finally,
section~\ref{section:conclusions} discusses the findings of this analysis, together with their impact on future
space-borne gravitational wave observatories. All the analysis presented in this work, has been performed using the 
dedicated data analysis toolbox, LTPDA~\cite{ltpda}.


\section{LPF dynamics
\label{section:dynamics}}

\begin{figure}[h!]
  \centering
  \includegraphics[width=0.9\linewidth]{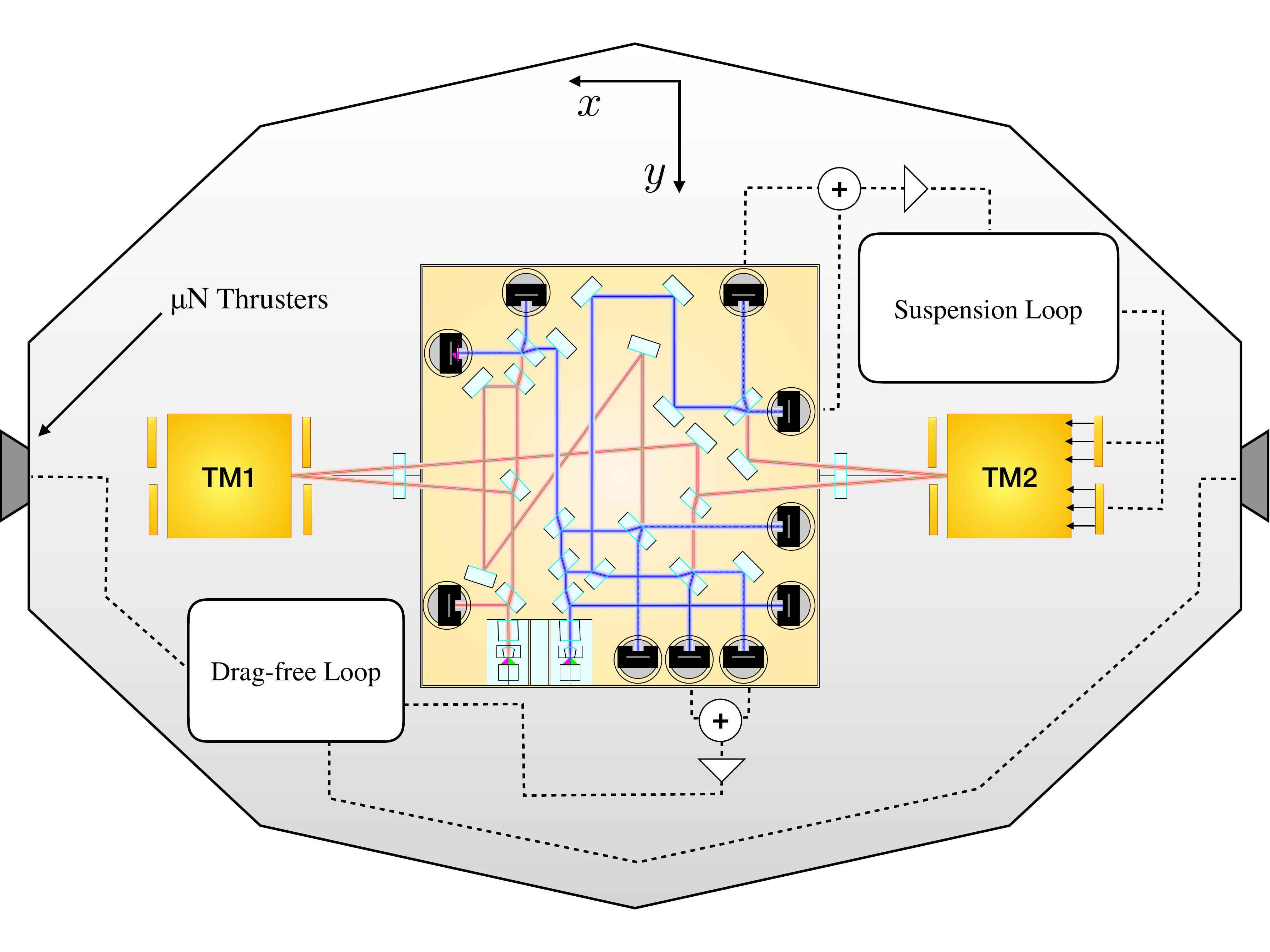}
\caption{A sketch of LPF emphasising the sensitive $x$-axis measurements. The two test-masses surrounded by the electrodes of their respective GRS, and separated by the Optical Bench, are shown inside the spacecraft. The $x_1[t]$ interferometer read-out is used to control the spacecraft to follow \tma via the {\em drag-free} control loop and by applying forces with the $\upmu$-Newton thrusters, and the electrodes are driven by the {\em electrostatic suspension} control loop to force \tmb to follow \tma.}
\label{fig:lpf}
\end{figure}

The main instrument on-board LPF, the LISA Technology Package~\cite{1742-6596-154-1-012007,0264-9381-22-10-001}, comprises two cubic test-masses and
their enclosures, and the Optical Metrology System (OMS)~\cite{oms4, oms3, oms2, oms1}. The test-masses and their surroundings form the
Gravitational Reference Sensor (GRS), which consists of the vacuum enclosure and the electrode housing~\cite{grs1}. For each test-mass, the electrode housing serves both as a capacitive position sensor in all 6 degrees-of-freedom, and as an
electrostatic force actuator. The OMS is responsible for the sensitive scientific measurement of the mission. It
measures $\Delta x [t] \equiv x_2 [t] - x_1 [t]$, the differential displacement between the two test-masses along
the joining $x$-axis, via means of heterodyne laser interferometry (see Fig.~\ref{fig:lpf} for details). 
The OMS also uses differential wavefront sensing to measure the differential angles between the two test masses
$\Delta \eta [t] \equiv \eta_2 [t] - \eta_1 [t]$ and
$\Delta \phi [t] \equiv \phi _2 [t] - \phi _1 [t]$.
The OMS has a second interferometer which 
measures $x_1$, the position of \tma relative to the spacecraft frame, $ \eta_1$ and $ \phi_1$ the  angles of the first test-mass relative to the spacecraft.
During the nominal and extension phase of operations the controllers operated the satellite in a specific drag-free scheme: driven by
the error signal of the $x_1$ interferometer, the spacecraft was commanded via the so-called {\em drag-free}
control loop to follow the first test-mass (or henceforth \tma) by applying forces with its $\upmu$-Newton thrusters.
At the same time, the second test-mass (\tmb) was electrostatically controlled via the so called {\em suspension loop} to
follow \tma  by keeping their distance fixed with a soft suspension of unity gain bandwidth near $1$~mHz. 

In this work, we define as calibration the determination of the purely dynamical parameters of the three body system, together with the 
instrumental parameters, mainly the gain coefficients of the capacitive actuators. We consider the OMS readout, $\Delta x [t]$, 
as our reference for the calibration. Considering the above, the equations of motion of the TMs and spacecraft allow the differential acceleration, $\Delta g_x$, to be calculated as~\cite{lpf_prl}
\begin{equation}
\label{eq:deltagfull}	
\Delta g_x [t] \equiv  \Delta\ddot{x} [t] - g_c[t] + \omega_2^2 \Delta x [t] + \Delta\omega^2_{12} x_1 [t]. 
\end{equation}
We introduce the term $\Delta g_x [t]$ to distinguish it from the differential acceleration corrected also for the inertial
forces, which we call $\Delta g [t]$, which is the final calibrated product (see below). The $(\,\dot{\,}\,)$ operator denotes the numerical time derivative, and $g_c [t]$ is
the commanded force per unit mass acting on \tma and \tmb respectively. It can be described as
\begin{align}
 \label{eq:fx2}
g_c [t] =& \lambda_2 \frac{F_{x_2}}{m_\text{TM$_2$}}[t - \tau_2] - \lambda_1 \frac{F_{x_1}}{m_\text{TM$_1$}}[t - \tau_1] =\nonumber\\
 =&\lambda_2 f_{x_2}[t - \tau_2] -  \lambda_1 f_{x_1}[t - \tau_1]
\end{align}
where $\lambda_1$ and $\lambda_2$ are the gain coefficients of the electrostatic commanded force on \tma and
\tmb, $m_\text{TM$_1$} =m_\text{TM$_2$} = 1.928\pm0.001$~kg is the mass of each TM, $f_{x_1}$ and
$f_{x_2}$ are the electrostatically applied forces $F_{x_1}$ and $F_{x_2}$ per unit mass, and $\tau_1$ and
$\tau_2$ the relative delay coefficients between the OMS and each of the GRS actuators. Different time delays   
between the various data processing units on-board are expected, but in the dynamical model of the system
it is the total relative delay between the two main read-out subsystems that is relevant. In the drag-free scheme, $F_{x_1}$ is nominally equal to zero, because no electrostatic forces are applied on \tma, the inertial reference along $x$.

In the environment of the spacecraft, a non-zero force gradient is present on both test masses. These force gradients can be attributed
mainly to gravity, electrostatic, and magnetic effects~\cite{ltpdefinitiondoc}. These gradients, or stiffnesses, produce forces acting
on the TMs in the presence of any relative motion, and enter eq.~(\ref{eq:deltagfull}) through the $\omega_j^2$ terms. $\omega^2_2$ is the stiffness
of \tmb~and $\Delta\omega^2_{12} = \omega^2_2 -\omega^2_1$ is the differential stiffness which couples the
spacecraft motion to the differential acceleration. 

Eq.~(\ref{eq:deltagfull}) describes a perfectly aligned system,
where the satellite is an inertial frame. However, these approximations were proved to be insufficient for the case of LPF. First of all, the unavoidable misalignment 
between the TMs, the GRS and the spacecraft, generates cross-coupling effects
that introduce cross talk signals in the main measurements. 
The main effect of these offsets is to couple spacecraft acceleration motion into the interferometer output, generating a
 ``bump'' in the higher frequency part of the  $\Delta g$ 
spectrum around 20 to 100~mHz~\cite{lpf_prl}. In~\cite{lpf_prl} we have subtracted this contribution by fitting a simple model which is a linear combination of the readouts
of the GRS and differential wavefront sensing. This model depends on the various geometrical configurations of the instrument and can be written as 
\begin{align}
\label{eq:bump}	
\delta g_\text{SC} [t] =	& b_1\ddot{\overline{\phi}}[t] +  b_2\ddot{\overline{\eta}}[t] + b_3\ddot{\overline{y}}[t] \nonumber \\
				&+ b_4\ddot{\overline{z}}[t] + b_5\overline{y}[t] + b_6\overline{z}[t]. 
\end{align}
where the $\left(\,\bar{\,}\,\right)$ denotes the mean displacement or rotation of both test-masses along the given coordinate $\in \{ \phi, \eta, y, z\}$. 
While this model performs quite well, it does not provide a solid physical interpretation of the various underlying cross-coupling effects, 
and therefore a more detailed study is necessary in order to better understand the data~\cite{wanner2017}. It is worth  
noting that this signal leakage into the sensitive differential measurement was only visible because of the much better than expected 
performance of the interferometers~\cite{lpf_prl}.

Secondly, LPF is a rotating reference frame, and its rotation introduces inertial forces with components along the
sensitive $x$-axis. We call the contribution of inertial forces on the differential acceleration, $g_\mathrm{rot}$. This term includes the
contribution from the centrifugal forces, $g_{\Omega}$, due to spacecraft angular velocity $\Omega$ with respect
to the J2000 reference frame, and the contribution due to the Euler forces, $g_{\dot{\Omega}}$, coming from a non zero
spacecraft rotational acceleration, $\dot{\Omega}$. The centrifugal term is always present when the angular velocity of the spacecraft is different from zero. On the contrary, the Euler forces appear only because of a geometrical offset of the TMs position with respect to their housing along $y$ and $z$ axes. It soon became very 
 clear during operations that these effects needed to be taken into account in the calculation of
$\Delta g$, which from eq.~(\ref{eq:deltagfull}) can now be written as 
\begin{equation}
\label{eq:deltagfull2}	
\Delta g [t] \equiv  \Delta g_x [t] + \delta g_\text{SC} [t] + g_\mathrm{rot} [t].
\end{equation}

\begin{figure*}
\begin{subfigure}{0.5\textwidth}
  \centering
  \includegraphics[width=0.9\linewidth]{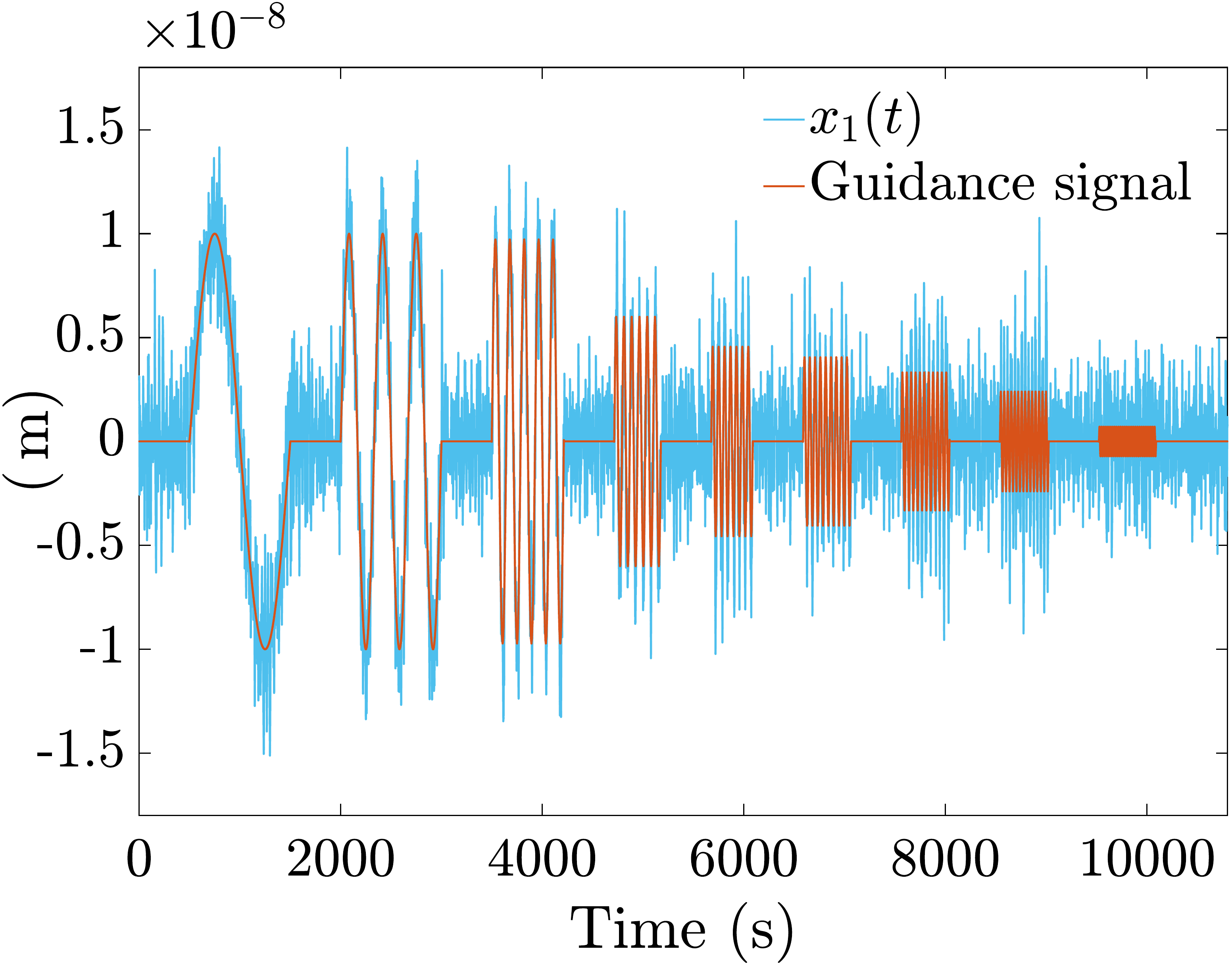}
  \caption{}
\end{subfigure}%
\begin{subfigure}{0.5\textwidth}
  \centering
  \includegraphics[width=0.9\linewidth]{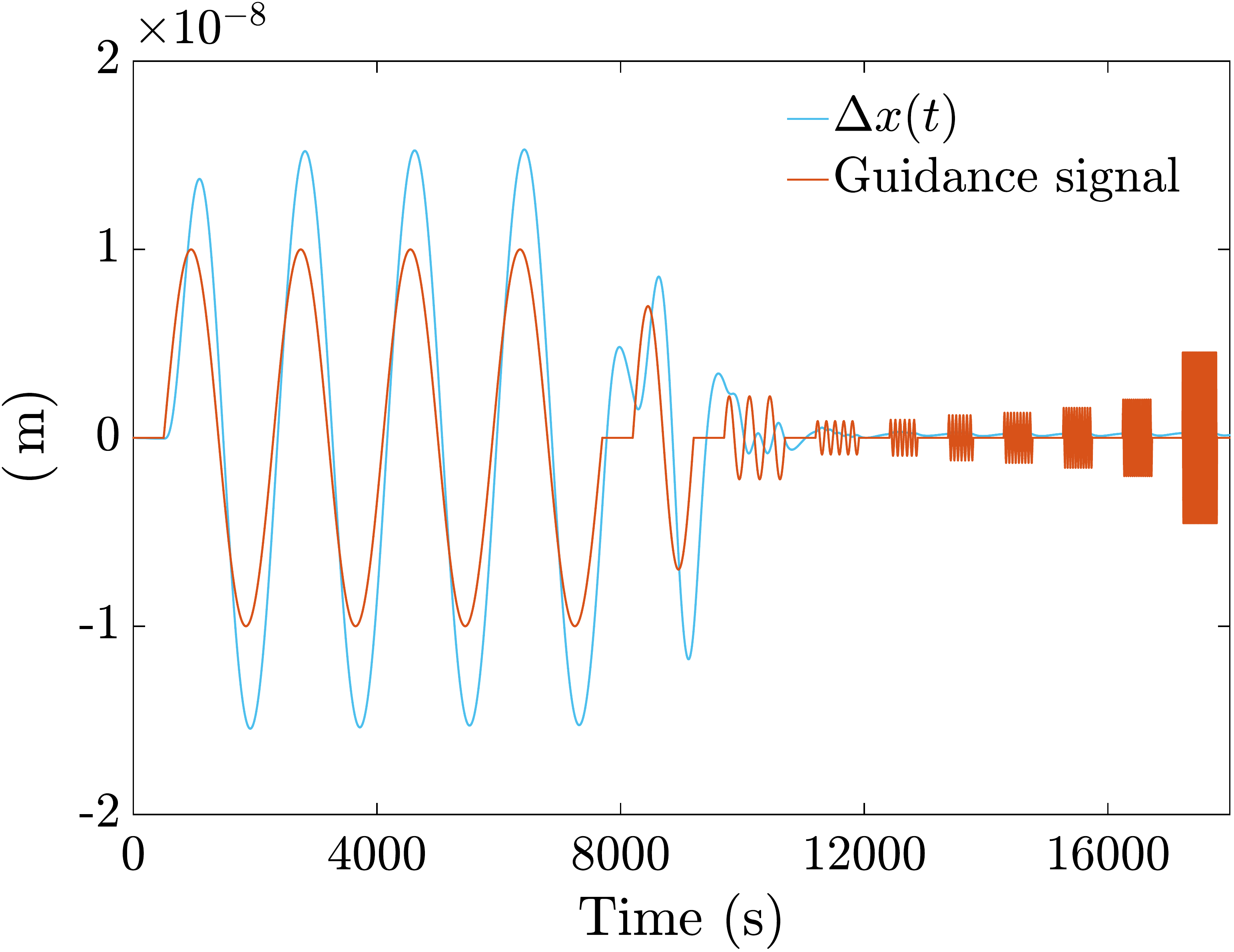}
  \caption{}
\end{subfigure}
\caption{Sequences of injections during the system identification experiment. {\it Left}: Injection of sinusoidal
	signals (red) into the drag-free loop. The response of the system, as recorded by the $x_1$ interferometer, is
	shown in light blue. {\it Right}: The same philosophy applies to the suspension loop. The measurement of
	the $\Delta x$ interferometer is again shown in light blue.}
\label{fig:injections}
\end{figure*}


\section{System Identification and Parameter Estimation
\label{section:sysid}}


In order to calibrate the dynamics of LPF described in the previous section, the so-called system identification experiments~\cite{PhysRevD.85.122004,
 1742-6596-363-1-012053, PhysRevD.89.062001} were regularly performed during the mission.
 Repeated experiments were necessary both to measure the long term stability of the system, and also because different working configurations 
 of the system and/or potentially different environmental conditions could yield different calibration parameters. These experiments consisted of sequences 
 of sinusoidal signal injections to various working points. For the large majority of the cases, we injected sinusoidal modulations into the sensitive $x$-axis at 
 frequencies between 1 and 50~mHz and amplitudes varying between 0.7 and 10~nm to the drag-free and suspension loops. The idea behind the design of the experiments 
  is to inject fake interferometric readouts, and let the system react to this apparent motion of the test-masses and spacecraft. In this way we excite 
  the dynamics of the system by effectively modulating the control position set-point, and induce high signal-to-noise ratio (SNR) for each of the dynamical 
  parameters to be estimated. Different injection frequencies 
  were used to break the degeneracies between force and readout effects, while at the same time, 
  the amplitudes of these injections were small enough to avoid the excitation of non-linearities of the system. The effect of the guidance 
  injections into $\Delta x$ and $x_1$ are shown in Figure~\ref{fig:injections}. 

During the course of the mission, different variations of the same experiments were defined and run on-board the satellite. This happened for two
 main reasons. The a posteriori knowledge of the lower than required levels of the acceleration noise, allowed us to inject signals with lower 
 amplitudes, while maintaining a sufficient SNR for the measurement of system parameters. Secondly, different flavors of the same principle 
 of experiment were performed in order to either target specific dynamical parameters, or to investigate the stability of the hardware. 
 Considering the complete set of calibration experiments, the injected frequency sweep in each channel ranged from 0.83 up to 53.3~mHz, while the maximum amplitude was always kept
  $\leq$1~$\mu$m. The total duration of the nominal injections was $6$~hours, while in special cases when investigating the system 
 stability, we injected single frequency and very low amplitude calibration tones into the suspension control loop that lasted $\simeq$65~hours.


\subsection{Fitting techniques/methodology}
\label{subsection:methodology}

The nature of the experiments, and the LPF mission in general, restricted the available analysis time during operations. The analysis team on duty 
had to fully analyse the data-sets within the time-span of two days, in order to retrieve the dynamical parameters and calibrate the $\Delta g [t]$ 
quantity. For that reason we adopted a fast and computationally light technique for the day-to-day analysis. First, we can combine 
eq.~(\ref{eq:deltagfull}) and (\ref{eq:fx2}) to rewrite them as
\begin{align}
	\label{eq:deltaglinear}
	\Delta g_x [t] \equiv 	& \Delta\ddot{x} [t] + \lambda_1 f_{x_1}[t] - \lambda_2 f_{x_2}[t] - C_1 \dot{f}_{x_1}[t]  \\ \nonumber 
					& + C_2 \dot{f}_{x_2}[t]  + \omega_2^2 \Delta x [t] + \Delta\omega^2_{12} x_1 [t],
\end{align}
where we have omitted the cross-coupling terms, and the inertial forces contributions $g_\Omega [t]$. Eq.~(\ref{eq:deltaglinear}) is a 
linearized form of the dynamics, where the $C_j$ parameter corresponds to the linearized delay coefficient multiplying the numerical first 
time derivative of the applied force per unit mass, $\dot{f}_{x_j}[t]$. The $C_j$ coefficients are equal to $C_j =  \tau_j \lambda_j$ if we approximate the calculated applied forces as
\begin{equation}
	\lambda_j f_{x_j}[t-\tau_j] = \lambda_j \left(f_{x_j}[t] - \tau_j \dot{f}_{x_j}[t] \right).
	\label{eq:delaylin} 
\end{equation}
Nominally, and for the majority of experiments performed, there were no commanded forces applied on \tma. 
But it was necessary to included it in eq.~(\ref{eq:deltaglinear}), because it was found that $f_{x_1}[t]$ was non-zero and non-negligible compared to the intrinsic noise level. The effective commanded force on \tma was due to imperfect digitization of the actuation waveforms used to apply a \tma torque~\cite{lambdamiscalibration}. Thus, two more free parameters had to be considered, the gain calibration coefficient $\lambda_1$, and its corresponding $C_1$ delay coefficient.

\begin{figure*}[!htb]
\begin{subfigure}{0.5\textwidth}
  \centering
  \includegraphics[width=0.9\linewidth]{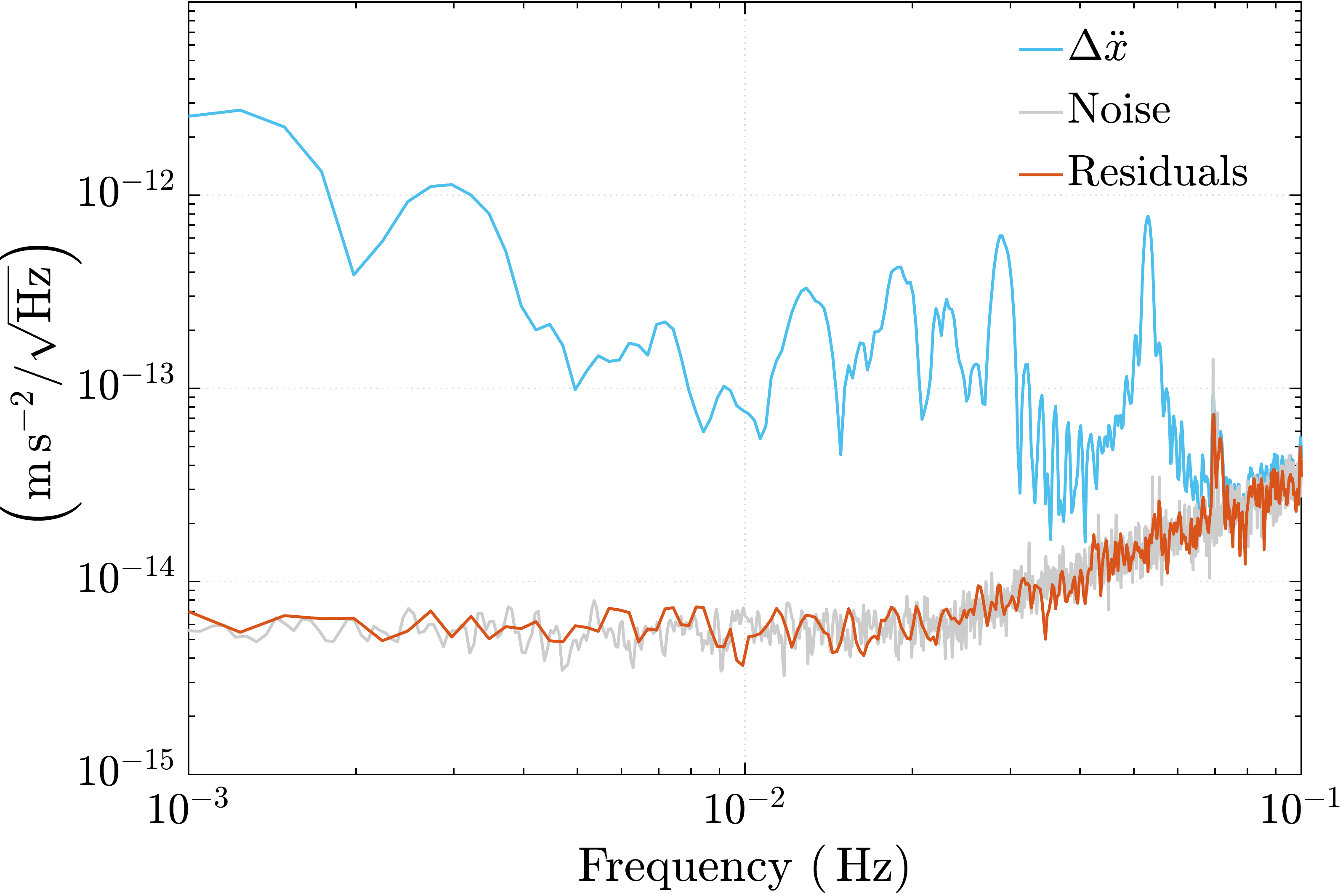}
  \caption{}
\end{subfigure}%
\begin{subfigure}{0.5\textwidth}
  \centering
  \includegraphics[width=0.9\linewidth]{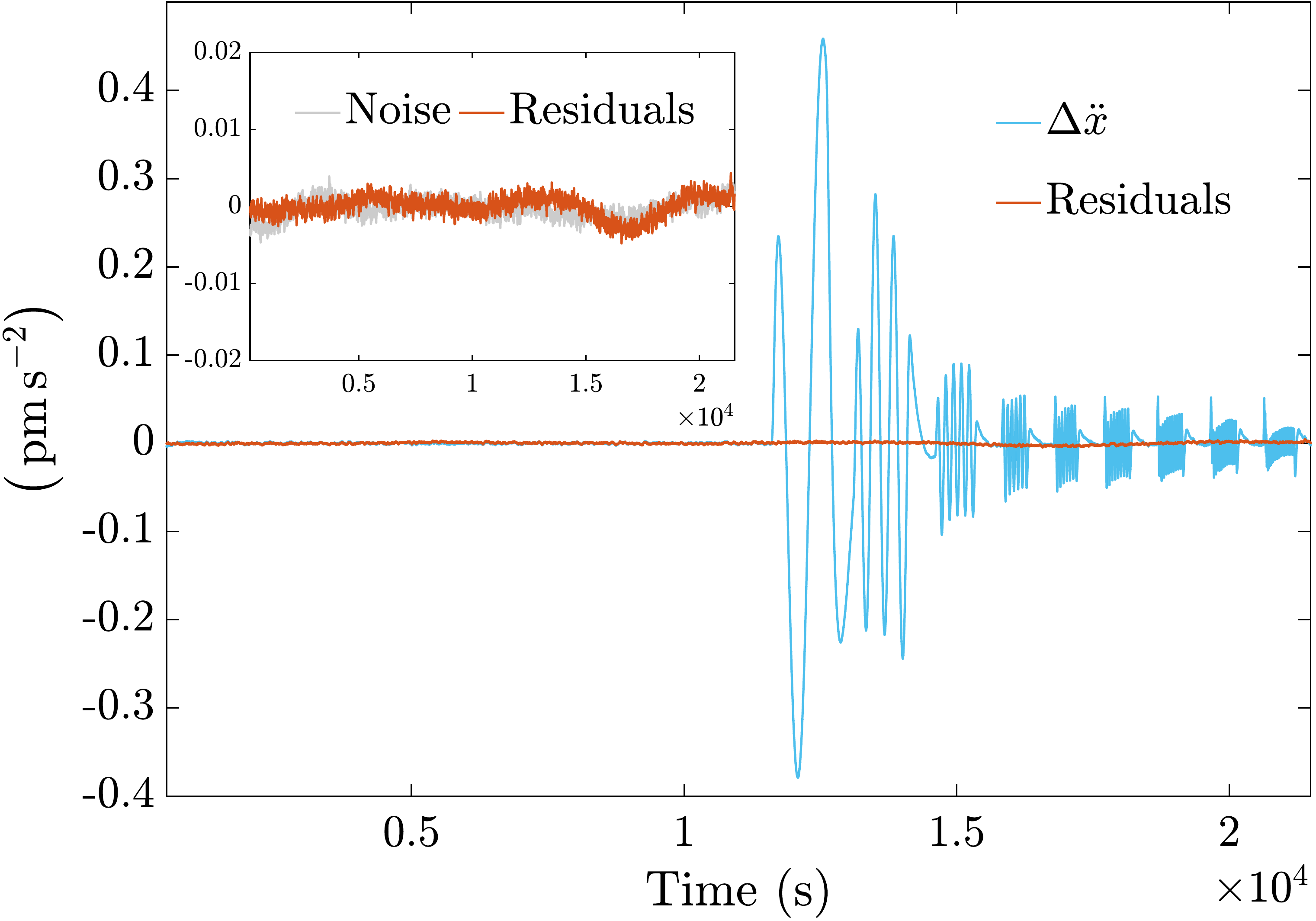}
  \caption{}
\end{subfigure}
\caption{{\em Left}: The spectra of the calculated residuals (in red) of the fit, compared to the induced signal in $\Delta\ddot{x} [t]$ (in light blue),
 and an acceleration noise measurement (in grey) performed the day prior the experiment. The particular data-set originates in the early stages of 
 the mission, around the 16$^\text{th}$ of April 2016. It is evident that the fit quality is good, since the residual levels match the spectrum of the 
 noise. {\em Right}: The comparison of time series of the signal induced in $\Delta\ddot{x} [t]$ with the calculated residuals after fitting the data. 
 In the embedded Figure, the comparison between the noise (grey) and the residuals (red) time series is shown. The time series have all been 
 low-pass filtered to aid comparison.}
\label{fig:residuals}
\end{figure*}

The model described above opens the possibility of adopting a suitable analysis scheme for operations, an Iterative Reweighted Least Squares (IRLS) 
algorithm \cite{stefanologarithmic}, where the problem reduces to solving a set of linear equations at each iteration. If we first identify the residuals
 from eq.~(\ref{eq:deltaglinear}) as $r(\param) \equiv \Delta g(t)$, then the IRLS procedure, at the  $n$-th iteration, can be written as
 \begin{equation}
	\chi^2_n = N_s \sum_{j \in Q} \frac{\overline{\abs{\tilde{r}_j (\param_n)}}}{\overline{\abs{\tilde{r}_j (\param_{n-1})}}}.
	\label{eq:irls}
\end{equation}
Here, $\tilde{r}$ represents the residuals in the frequency domain at each frequency bin $j$, $\param$ the given parameter set to be estimated, and $N_s$ the number of 
data stretches used to perform the averaging for the computation of the Power Spectral Density (PSD) of the signals. For each of the 
experiments this technique yielded a good quality of fit, with the calculated residuals being 
compatible with the noise measurements (see Figure~\ref{fig:residuals} and section \ref{section:results}). 


\subsection{Analysis of the complete dataset of system identification experiments}
\label{subsection:jointanalysis}

The system parameters of the gain actuation $\lambda_j$ and the delay coefficient $C_j$ from eq.~(\ref{eq:deltagfull}) for each TM can be considered as constants for the duration of 
the mission. This assumption was based on the stationarity of the instrument, and it was proven to be true a posteriori, from the results of the 
calibration experiments (see sections~\ref{subsection:methodology} and~\ref{section:results}). 

The dynamical parameters of the stiffnesses 
$\omega_{j}^2$ however, depended mostly on the actuation configuration and the gravitational balance of the instrument. If the gravity gradient 
of the satellite can be assumed constant, and we can model the dependance of the stiffness on the electrostatic actuation, then in principle it 
is possible to adopt a joint fitting scheme, where all the calibration experiments can be analyzed together. This will allow us to accurately estimate the background stiffness $\omega^2_{j, \, 0}$, which is expected to be dominated by gravitational effects~\cite{0264-9381-33-23-235015} and by electrostatic stiffness from the 100 kHz capacitive sensing bias~\cite{SPIE}.  Additional stiffness from TM electrostatic charge~\cite{SPIE} is negative and smaller than our experimental resolution, calculated to range between $0$ and $-3 \times 10^{-9}~\text{s}^{-2}$ for the $\pm ~3~\text{pC}$ of TM charge present during science operations.  Likewise, gravitational stiffness variations from cold gas fuel depletion~\cite{0264-9381-33-23-235015} are calculated to be below $10^{-9}~\text{s}^{-2}$ over the course of the mission.  For these reasons, we treat the background stiffness as a constant in our analysis.

LPF operates in the so-called ``constant stiffness" actuation configuration by which the commanded force $g_c$ can be applied within a range of force and torque authority (maximum force and torque applicable) while keeping the electrostatic stiffness value constant. This constant value depends only on a linear combination of the force and torque authority. A given relative fluctuation in an applied actuation voltage on one electrode produces a force fluctuation that is proportional to the magnitude of applied force from that electrode.  As such, a higher level of force and torque authority corresponds to a higher force noise induced by the actuation subsystem because of in-band amplitude voltage fluctuations~\cite{lpf1}.
Throughout the mission, different levels of force and torque authority were introduced. 
For example, the so-called ``nominal actuation configuration'' sets the maximum force acting on \tma along $x$ to $F_{{\rm max}, \,1} = 0$, 
and on \tmb to $F_{{\rm max}, \,2} = 2200$~pN, while the maximum applied torques along $\phi$ (around the $z$-axis) were set to 
$N_{{\rm max}, \, 1} = N_{{\rm max}, \, 2} = 10.4$~pNm. This authority level was set at the beginning of the science operations as a safe one, based on the foreseen level of static gravity field that needed to be compensated by electrostatic actuation. However, given the level of gravitational balancing measured on board, which was consistently below $50~\text{pN}$ throughout the entire science phase of the mission~\cite{lpf_prl,lpf_prl2}, the authority level was gradually reduced to reach the values of $\{F_{{\rm max}, \,1} = 0$, $F_{{\rm max}, \,2} = 50$~pN, $N_{{\rm max}, \, 1} = 1.5$~pNm, $N_{{\rm max}, \, 2} = 1~\text{pNm}\}$, reducing in turn the actuation noise. 
Therefore, the electrostatic component of the stiffness should depend 
on the constant stiffness actuation configuration, and on the $x$ and $\phi$ maximum actuation authorities, but not on the time-variable applied 
force and torque commands. Considering the above, a simplified model for the stiffness can be defined as:
\begin{equation}
	\omega_{j, \, \text{tot}}^2  = \omega_{j, \, 0}^2 + \alpha_{x_j}F_{{\rm max}, \,j} + \alpha_{\phi_j}N_{{\rm max}, \, j}.
	\label{eq:stiff_model}
\end{equation}
The $\alpha_{x_j}$ coefficient couples to the $x$ force authority to give $x$ stiffness component, and can be written as \cite{nico_lisa_symp}
\begin{equation}
	\alpha_{x_j} = -\lambda_V^2\frac{1}{m_{\text{TM}_j}}\frac{\frac{\partial^2 C^\ast_X}{\partial x^2}}{\frac{\partial C^\ast_X}{\partial x}},
	\label{eq:alphax}
\end{equation}
where $C^\ast_X \equiv C_X + C_{X,\,h}$ represents the total X electrode capacitance, 
 the first contribution being the capacitance from electrode to TM and the second capacitance 
from the electrode to the grounded guard ring surfaces. 
$\lambda_V$ represents a nominal miscalibration factor of the nominal applied voltages, calculated to be $\lambda_V = 1.066 \pm 0.002$ from a known voltage reference mismatch in the actuation feedback circuitry. 
The $\alpha_{\phi_j}$ coefficient couples to the $\phi$ force authority to give the stiffness along the $x$-axis, and should be given by
\begin{equation}
	\alpha_{\phi_j} = -\lambda_V^2\frac{1}{m_{\text{TM}_j}}\frac{\frac{\partial^2 C^\ast_X}{\partial x^2} - 4\frac{\left(\frac{\partial C^\ast_X}{\partial x}\right)^2}{C_\text{tot}}}{\frac{\partial C^\ast_X}{\partial \phi}}.
	\label{eq:alphaphi}
\end{equation}
The expected values for the $\alpha_{x_j}$ and $\alpha_{\phi_j}$ coefficients were calculated from eq.~(\ref{eq:alphax}) and (\ref{eq:alphaphi}) 
to be around $-320$~kg$^{-1}$~m$^{-1}$ and $-26500$~kg$^{-1}$~m$^{-2}$ respectively~\cite{nico_lisa_symp}. 

The above simplified and global model of the stiffnesses of the system would, in principle, allow for a joint-fit analysis of the complete 
data-set of all system identification experiments performed over the entire duration of the mission. Thus, the background stiffness 
$\omega_{j, \, 0}^2$, which includes all the effects other than the electrostatic ones, can be considered as a free parameter, with a 
common value for all the experiments from the beginning to the end of the mission. 

At the same time, the parameters depending on the geometrical orientation of the three bodies of the system need to be taken into account. 
One such parameter is the $\delta_\text{ifo}$ which describes the signal leakage from the $x_1 [t]$ measurement to $\Delta x [t]$.
 Although this effect is classified as a read-out cross-coupling and not as a dynamical parameter, it is necessary to include it in the 
 fit in order to subtract all the signal induced, and perform goodness-of-fit tests to the resulting residuals. It is worth to mention that
  non-negligible correlation between the $\delta_\text{ifo}$ and the $b_i$ parameters appearing in eq.~(\ref{eq:bump}) is expected. 
  The effect described from $\delta_\text{ifo}$ is associated to the motion of the SC with respect to the optical bench, and this relation is 
  currently under study~\cite{wanner2017}. 
 The $\delta_\text{ifo}$ parameter could take different values when the TMs are re-aligned, intentionally or not. Such re-alignments can occur due to grabbing and re-releasing 
 the TMs after dropping to safe mode of operation, or due to considerable changes of 
 temperature gradients inside the satellite environment. We then have $\delta_\text{ifo} \equiv \delta_{\text{ifo}, \, {\mathrm k}}$, 
 where ${\mathrm k}$ refers to the different ``geometrical orientation epochs'' of the three-body system. We can then rewrite 
 eq.~(\ref{eq:deltaglinear}) in a very similar way as 
\begin{align}
	\label{eq:deltagjointfit}
\Delta g_x [t] = 	& \Delta\ddot{x} [t] - \lambda_2 f_{x_2}[t] + \lambda_1 f_{x_1}[t] \\ \nonumber
              		& + \omega_{2, \, \text{tot}}^2 \Delta x[t] + \Delta\omega_\text{tot}^2 x_1 [t] \\ \nonumber
          		& - C_1 \dot{f}_{x_1}[t] + C_2 \dot{f}_{x_2}[t] - \delta_{\text{ifo}, \, {\mathrm k}} \ddot{x}_1 [t].
\end{align}
where
\begin{align}
	\Delta\omega_\text{tot}^2  =  	& \omega_{2, \, \text{tot}}^2 - \omega_{1, \, \text{tot}}^2 \\ \nonumber
						=	& \omega_{2, \, 0}^2 + \alpha_{x_2}F_{\text{max}, \,2} + \alpha_{\phi_2}N_{\text{max}, \, 2} \\ \nonumber
          				  		& - \left(\omega_{1, \, 0}^2 + \alpha_{x_1}F_{\text{max}, \,1} + \alpha_{\phi_1}N_{\text{max}, \, 1}\right).
	\label{eq:deltastiff_model}
\end{align}
In eq.~(\ref{eq:deltagjointfit}) we have assumed common $\lambda_j$, $C_j$, and $\omega_{j, \, 0}^2$ for all the experiments, while only the orientation 
depended $\delta_{\text{ifo}, \, {\mathrm k}}$ varies through the mission duration.
After comparing the various events that caused rotation and displacement of the TMs, we concluded that ${\mathrm k} = \{1,2,3,4\}$, adding three
 additional parameters to be estimated. It is worth mentioning that, in order to verify this procedure, we have repeated the analysis described here for a smaller set of ${\mathrm m}$ experiments, by assuming different $\lambda_{j, \, {\mathrm m}}$ and $C_{j, \, {\mathrm m}}$ coefficients, which yielded very similar results. 
 
 In total, we averaged over 13 of the 22 available experiments of this kind, over the $\sim2$ years duration of the nominal and extension operations of the mission. We chose to exclude from the fit the experiments that were too short ($<6~\mathrm{h}$) to be sensitive enough to the stiffnesses parameters.   

\begin{figure*}[!htb]
  \centering
  \includegraphics[width=0.9\linewidth]{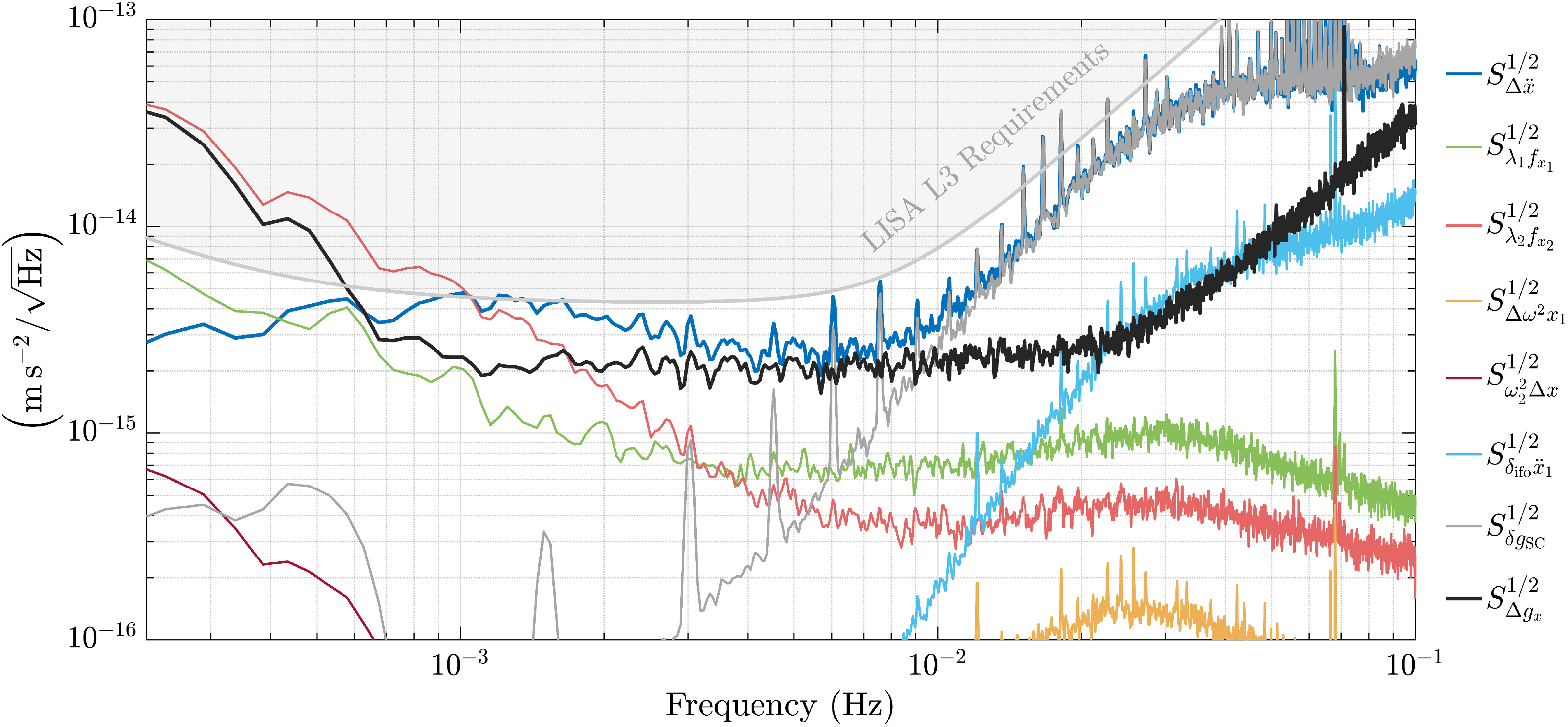} 
\caption{Spectra of the breakdown of the differential acceleration noise between the two TMs. Starting from the raw measurement of $\Delta \ddot{x} [t]$ (blue), we calculate $\Delta g_x[t]$ (black) from eq.~(\ref{eq:deltagfull2}) using the parameters from Table~\ref{tab:jointfit}. The data plotted here refer to a noise measurement performed between 17/05/2017 and 23/05/2017, while the ASD was computed by assuming 35 data stretches overlapped by 50\%. The applied force on \tma along $x$ (green curve) is nominally zero, however for this particular run compensating voltages acting on the electrodes around \tma, caused a non-zero $x$ force. The cross-coupling contribution originating from the SC jitter is also shown for comparison (grey). See text for details.}
\label{fig:breakdown}
\end{figure*}

Finally, the time-dependencies of the levels of the noise need to be considered. For example, the Brownian noise contribution, visible at frequencies from 1~mHz to $\sim$30~mHz, was observed to decrease with time~\cite{lpf_prl}. For that reason, when forming the noise weighted inner product between two real time
  series $a$ and $b$~\cite{PhysRevD.49.2658}
\begin{equation}
	\left( a | b \right) = 2 \int\limits_0^\infty \mathrm{d}f \left[ \tilde{a}^\ast(f) \tilde{b}(f) + \tilde{a}(f) \tilde{b}^\ast(f) \right]/\tilde{S}_n(f)\,,
	 \label{eq:inprod}
\end{equation}
that enters the gaussian likelihood as
\begin{equation}
	\pi(y|\param) = C \times e^{ -\textstyle{1\over 2}\big( r(\param) \big| r(\param) \big)} = C \times e^{-\chi^2/2},
	\label{eq:llhgaussian}
\end{equation}
the power spectrum of the noise $S_n(f)$ needs to be calculated from a noise measurement with the system in the same configuration, and close in time, to the relevant calibration experiment $n$. Considering all the above, since the different experiments are independent, and without any overlap, we can approach the
  parameter estimation part of the analysis by assuming a joint likelihood scheme \cite{PhysRevD.82.122002}, defined as
\begin{equation}
 	\Lambda_\text{tot}(\Theta) = \sum_i^{N_\text{exp}} \mathrm{log} \left( \pi(y_i | \param_i, \mathcal{M}_i) \right),
	\label{eq:jointlh}
\end{equation}
where $y_i$ is the data-set of the $i$-th experiment, and $\param_i \in \Theta$ the corresponding parameter set, with $\Theta$ being the complete 
parameter set. The $\pi(y_i | \param_i, \mathcal{M}_i)$ is the corresponding likelihood function for the given data-set $y_i$ and model $\mathcal{M}_i$. 
The remaining caveat to consider is the possible non-stationarities of the noise for each of the $i$ experiments. Presence of spurious data, such as 
glitches or noise bursts, could cause biased estimation of the parameters. This is addressed if we adopt a likelihood function with longer tails, 
to properly account for the uncertainty of the noise model of each of the experiments, such as in~\cite{rover1} and~\cite{rover2}. With the proper 
assumptions we can also utilize a marginalized likelihood formulation where all the noise FFT coefficients are marginalized out of the expression, 
like in~\cite{stefanologarithmic}. Even if the data in most cases were of high quality, without any outliers~\cite{lpf_prl}, we have sampled the 
posterior distributions for all cases with Markov Chain Monte Carlo algorithms~\cite{PhysRevD.82.122002}, and found that all 
implementations yielded consistent results within the one-$\sigma$ margin. With the estimated parameters (see Table~\ref{tab:jointfit}, and section~\ref{section:results} for details), 
we were able to subtract all the signal power induced, yielding a satisfactory fit with the residuals being compatible with the noise for all of 
the experiments analysed. The contribution of each term appearing in eq.~(\ref{eq:deltagjointfit}) is shown in Figure~\ref{fig:breakdown}, together with the final calibrated $\Delta g_x [t]$ product. 

\begin{table}[h!]
\small
\begin{center}
\begin{tabular}{| l r@{\hd$\pm$\hd}l |} 
\hline
Parameter & \multicolumn{2}{c|}{Estimated $\pm \sigma$}   \\
\hline \hline
{ $\lambda_1$ }			 							&  $ 1.0748$	& $0.0001$  	\\
{ $\lambda_2$ }			 							&  $ 1.0776$ 	& $0.0001$  	\\
{ $\omega_{1, \, 0}^2$~(s$^{-2}$)}						&  $-(3.99$ 	& $ 0.07)\times10^{-7}$ 	\\
{ $\omega_{2, \, 0}^2$~(s$^{-2}$)}						&  $-(4.19$ 	& $ 0.04)\times10^{-7}$ 	\\
{ $C_1$~(s) }										&  $0.149$ 	& $ 0.004 $  			\\
{ $C_2$~(s) }										&  $0.186$ 	& $ 0.004 $  			\\
{ $\alpha_{x_1}$~(kg$^{-1}$~m$^{-1}$) } 					&  $-345$ 		& $ 9$  	\\
{ $\alpha_{\phi_1}$~(kg$^{-1}$~m$^{-2}$$\times10^{3}$) }		&  $-28.2$		& $ 0.9$  	\\
{ $\alpha_{x_2}$~(kg$^{-1}$~m$^{-1}$)}					&  $-317$		& $ 6$  	\\
{ $\alpha_{\phi_2}$~(kg$^{-1}$~m$^{-2}$$\times10^{3}$)  } 	&  $-27.4$		& $ 0.8$  	\\
{ $\delta_{\rm ifo, \,1}$ }								&  $(-2.2$ 		& $ 0.2)\times10^{-5}$	\\
{ $\delta_{\rm ifo, \,2}$ }								&  $(-1.40$ 	& $ 0.06)\times10^{-5}$	\\
{ $\delta_{\rm ifo, \,3}$ }								&  $(-2.09$ 	& $ 0.06)\times10^{-5}$	\\
{ $\delta_{\rm ifo, \,4}$ }								&  $(1.91$ 	& $ 0.08)\times10^{-5}$	\\
\hline 
\end{tabular}
\caption[Parameter estimation results for the joint fit assuming common gain for each experiment]{Parameter estimation results for the joint fit scheme,
 assuming common gain coefficients $\lambda_j$ and background stiffnesses $\omega_{j, \, 0}^2$ for each of the system calibration experiments considered in the fit (see
  section~\ref{section:results} for a detailed explanation).}
\label{tab:jointfit}
\end{center}
\end{table}                     


\section{Inertial Correction\label{section:inertial}}

During the LPF science operations, it was soon recognized that $S_{\Delta g_x}^{1/2}$, the ASD  of $\Delta g_x[t]$ as
calculated in the previous section, exhibited a noise fluctuation below $0.5~\mathrm{mHz}$ which exceeded the
contribution from the expected dominant actuation noise. In order to understand the origin of this noise excess, it
is important to point out that the estimation of $\Delta g_x$ in eq.~(\ref{eq:deltagjointfit}) produces
a time-series which reintroduces all the external forces acting on the TMs at low frequencies that
were suppressed by the action of the electrostatic suspension loop. 
As already stated in section~\ref{section:dynamics}, inertial forces due to the noisy rotation of the spacecraft act on
the TMs and the effect of these forces is expected to appear below $0.5~\mathrm{mHz}$.

The appearance of the inertial forces is directly linked to the fixed orbital attitude maintained by the spacecraft
in order to keep the solar panel pointing towards the Sun and the communication antenna towards the Earth. 
The spacecraft attitude is maintained at very low frequency by means of the error signal 
of an Autonomous Star Tracker (AST) which tracks the movement of the spacecraft with respect
to the fixed stars in the camera field. This error signal indirectly feeds, through the so called {\em attitude control loop},
the $\upmu$-Newton thrusters that command the spacecraft to
slowly rotate with respect to the fixed stars. An instrumental read-out noise of the AST could produce a noisy rotation of the spacecraft, which in turn makes the inertial forces appearing in the relative acceleration signal between the TMs, $\Delta g_x$.

\subsection{The inertial contribution to $\Delta g_x$}
\label{subsection:contribution}

In a rotating frame, three apparent force contributions on a test body can arise: the centrifugal force,
the force associated to the Coriolis effect and the Euler force. In the case of LPF the designed control scheme ensures that  
when no excitation signals are being injected there is no appreciable relative velocity between spacecraft and TMs. For this reason the Coriolis effect does 
not significantly affect the data. However, both centrifugal forces and Euler forces appeared to contribute significantly to the very low frequency ASD of the acceleration
noise.

In the case of the LPF three-body system, the acceleration of one TM due to inertial forces and measured by a rotating reference
frame attached to the spacecraft is given by
\begin{equation}\label{eq:inertialacc}
\ddot{x}[t] =  (\dot{\vec{\Omega}}[t] \times \vec{r} + \vec{\Omega}[t] \times (\vec{\Omega}[t] \times \vec{r})) \cdot \hat{x}
\end{equation}
where $\vec{\Omega}$ is the spacecraft angular velocity with respect to J2000 reference frame 
and $\vec{r}$ is the position vector of one TM with respect to the reference frame attached 
to LPF with the origin in the center of the optical bench. For the sake of clarity, from now on the time dependency of the angular velocity and acceleration will 
not be shown explicitly unless there is a strong need to show this dependence.
For each TM, the only contribution to $\Delta g_x$ is that coming from the $\hat{x}$ component
of the acceleration as in equation (\ref{eq:inertialacc}). Assuming the direction from the center of the optical bench towards \tma as the positive direction in the chosen refence system, it
is possible to write for \tma and \tmb, respectively,
\begin{align}
\ddot{x}_1[t] =&  -(\Omega^2_{\phi} + \Omega^2_{\eta}) x_1 + (-\dot{\Omega}_{\phi}+ \Omega_{\eta}\Omega_{\theta}) y_1 +\\\nonumber
&+(\dot{\Omega}_{\eta} + \Omega_{\phi}\Omega_{\theta}) z_1 \\
\ddot{x}_2[t] = & -(\Omega^2_{\phi} + \Omega^2_{\eta}) x_2 + (-\dot{\Omega}_{\phi}+ \Omega_{\eta}\Omega_{\theta}) y_2 +\\\nonumber
&+(\dot{\Omega}_{\eta} +\Omega_{\phi}\Omega_{\theta}) z_2\nonumber
\end{align}
where $\{x_1, y_1, z_1\}$ and $\{x_2, y_2, z_2\}$ represent the coordinates of 
\tma and \tmb in the defined reference system, and $\Omega_{\phi}$, $\Omega_{\eta}$ and $\Omega_{\theta}$
are the angular velocities around the axes $z$, $y$ and $x$, respectively.

The contribution of the inertial forces to $\Delta g_x$ can be written as the differential acceleration
between the two TMs
\begin{align}
g_\mathrm{rot}[t] = \ddot{x}_{2} - \ddot{x}_{1}  = & -(\Omega^2_{\phi} + \Omega^2_{\eta}) (x_2 - x_1) +\\\nonumber &(-\dot{\Omega}_{\phi} + \Omega_{\eta}\Omega_{\theta}) (y_2 - y_1) +\\ \nonumber
&(+\dot{\Omega}_{\eta} + \Omega_{\phi}\Omega_{\theta}) (z_2 - z_1).
\label{eq:x12}
\end{align}
In the case of perfect alignment and centering between the axis joining the center of the two TMs and the optical axis
of the interferometer, $x_2 - x_1$ is equal to $0.376$~m, that is the distance, $d$, between
the two TMs, a known quantity by design and on-ground measurements. Moreover, $y_2 - y_1 = z_2 - z_1 = 0$~m. This would reduce the
acceleration noise contribution due to inertial forces to
\begin{equation}
g_\mathrm{rot}[t]  = -(\Omega^2_{\phi} + \Omega^2_{\eta}) d =g_{\Omega}[t],\label{eq:centrifugal}
\end{equation}
that is just the centrifugal contribution. 

Any misalignment between the axis joining the two TMs and the interferometric
$x$-axis will still give
$x_2 - x_1$  equal to $d$, but we will now have $y_2 - y_1$  equal to $\delta y$ and $z_2 - z_1$ to
$\delta z$, that is the relative offsets of the two TMs along $y$ and $z$.
A new equation for the inertial forces contributing to $\Delta g_x$ can be written as
\begin{align}
g_\mathrm{rot}[t] = & -(\Omega^2_{\phi} + \Omega^2_{\eta}) d  -\dot{\Omega}_{\phi} \delta y +\dot{\Omega}_{\eta}\delta z +\\\nonumber 
& + \Omega_{\eta}\Omega_{\theta} \delta y  + \Omega_{\phi}\Omega_{\theta}\delta z\label{eq:gomega}
\end{align}
The terms of order $\Omega^2$ compose the centrifugal contribution, while those proportional to
$\dot{\Omega}$ form the Euler force contribution.
The noise in the angular velocity, $\Omega$, is typically of the order of $10^{-7}~\mathrm{rad}~\mathrm{s}^{-1}/\sqrt{\mathrm{Hz}}$, while
$\delta y$ and $\delta z$ are expected to be of the order of $10^{-5}-10^{-6}~\mathrm{m}$. Thus any
misalignment of the TMs which multiplies $\Omega^2$ can be neglected to the first order in $\Delta g$.
However, the terms coming from the Euler force, that is proportional to the angular acceleration, which
in turn can reach also $10^{-10}-10^{-11}~\mathrm{rad}~\mathrm{s}^{-2}/\sqrt{\mathrm{Hz}}$, are
not negligible to the first order. This has been verified by on-board measurements.

The final calibrated $\Delta g[t]$ in the band $0.1 - 20$~mHz can be calculated by subtracting from
equation (\ref{eq:deltagjointfit}) the first order expression for $g_\mathrm{rot}$ given by
\begin{align}
g_\mathrm{rot}[t] = &g_{\Omega}[t] + g_{\dot{\Omega}}[t] = \\\nonumber 
=& -(\Omega^2_{\phi}[t] + \Omega^2_{\eta}[t]) d -\dot{\Omega}_{\phi}[t]  \delta y +\dot{\Omega}_{\eta}[t]  \delta z \label{eq:gomegafinal}
\end{align}

\subsection{The sub mHz $\Delta g_x$ calibration}
\label{subsection:calculation}

As already stated in the previous subsection, the only known quantity in equation (\ref{eq:gomega}) is $d$, the distance between the two TMs. A measurement of the spacecraft angular velocity $\Omega$ is available on-board LPF through the AST quaternions
time-series. However, at sub-mHz frequencies the relative measured angular velocity is too noisy. 
Any direct use of this
quantity to correct $\Delta g_x$ through equation (\ref{eq:gomega}), would make the AST read-out noise dominate. A
way to calculate $\Omega$ that is not affected by AST noise is to combine different measurements of the
angular velocity for different frequency bands. For instance, the sub-mHz fluctuating part of the spacecraft angular
velocity can be calculated using
the applied electrostatic torques on the TMs along $\phi$, $\eta$ and $\theta$.
Along these degrees of freedom, the TMs are electrostatically controlled to follow the rotation of the spacecraft.
This electrostatic control is driven by the angular rotation of the TMs measured by the on-board interferometers through
the differential wavefront sensing read out. The angles are measured with respect to a reference attached to the optical
bench, which in turn is rigidly attached to the spacecraft. It is then a very precise measurement of the time variation of the angular relative position of the TMs and the spacecraft, an in-loop quantity, which is translated into an applied torque.
Then, from the applied torque, which is available from telemetry, it is possible to recover by time
integration and high pass filters the fluctuating part of the spacecraft angular rotation, $\Omega_\mathrm{noise}$.
It is also possible to calculate the DC part of the angular velocity, $\Omega_\mathrm{DC}$, by lowpass filtering the angular velocity 
as measured by the quaternions and fitting it to a polynomial to remove the noise due to the read out. All together, $\Omega = \Omega_\mathrm{noise} + \Omega_\mathrm{DC}$ gives an
estimation of the spacecraft angular velocity which is free of the AST read out noise. With this signal the subtraction of the
centrifugal force term given by $-(\Omega^2_{\phi}[t] + \Omega^2_{\eta}[t]) d$ from $\Delta g_x$ can be performed directly.

The other quantity that needs to be measured is the angular acceleration, $\dot{\Omega}$. In this case, the applied
torque to the TMs is already a good estimation of the angular acceleration. However, in this
case, neither $\delta y$ nor $\delta z$ are known, and a fit to recover them is necessary. The analysis is performed in the
frequency domain using the technique explained in section~\ref{subsection:methodology}, by fitting the Euler contribution to $\Delta g_x + g_\Omega$, which is the calibrated differential acceleration noise of eq.~(\ref{eq:deltagjointfit}) already corrected for the centrifugal contribution. The Euler force term can be
rewritten in terms of applied torques on the TMs as 
\begin{equation}
g_{\dot{\Omega}}[t] = -\dot{\Omega}_{\phi}  \delta y +\dot{\Omega}_{\eta} \delta z = d\left(-\frac{N_{\phi}}{I_{zz}}\delta \phi + \frac{N_{\eta}}{I_{yy}}\delta \eta\right)
\end{equation}
where $N_{\phi}$ and $N_{\eta}$ are the applied torques around $z$ and $y$, $I_{zz}$ and $I_{yy}$ the respective
TMs momenta of inertia and $\delta \phi = \delta y /d$ and $\delta \eta = \delta z /d$ the corresponding
misalignment angles.
$\delta \phi$ and $\delta \eta$ are the free parameters of the fit.
It was discovered that for three different periods of operations, three different misalignment values of 
the optical axis and the TMs joining axis must be considered. The change of the parameters corresponds to particular manoeuvres on the spacecraft which physically changed the position of the TMs. A global fit of the Euler force term to $\Delta g[t] + g_\Omega[t]$ has been performed, using for each period data from different noise stretches. In this way a statistically valid  estimation of the parameters value has been recovered for the three periods assumed (see Table \ref{tab:Eulerfit}).
These values are then used to directly subtract the effect of angular acceleration from all the acceleration noise data during the LPF mission.
It is important to point out that the effect of the Euler force in $\Delta g_x(t)$ is observationally indistinguishable from that of the actuation crosstalk between the applied torque on the test mass and the applied force along the sensitive $x$-axis. The origin of such a crosstalk has to be seen in the possible imperfections of the commanded voltages on the single electrodes which transform a pure torque signal in a torque plus a small force. The effect is in principle present for any rotational degree of freedom but is expected to be more pronounced for the torque around $\phi$ because of the geometry of the system. Indeed, the same electrodes are used to apply a force on $x$ and a torque around $\phi$. However, the observation that the roll of the spacecraft is very high when the subtraction of the Euler force from $\Delta g_x + g_{\Omega}$ is bigger points to the fact that this last effect is by far the dominant one.

\begin{table}[h!]
	\small
	\begin{center}
		\begin{tabular}{| c c c c |} 
			\hline 
							& \multicolumn{3}{c|}{Estimated $\pm \sigma$} \\
							&\multicolumn{1}{c}{before} &\multicolumn{1}{c}{from}  &\multicolumn{1}{c|}{after} \\
			Parameter &\multicolumn{1}{c}{19 June 2016} &\multicolumn{1}{c}{19 to 25 June 2016} &\multicolumn{1}{c|}{25 June 2016} \\
			\hline \hline
			{ $\delta \phi$~(mrad) }		&  $ -0.47\pm 0.03 $ &  $ -0.40\pm 0.03 $ &  $ -0.39\pm 0.02 $\\
			{ $\delta \eta$~(mrad)  }		&  $ -0.066\pm 0.007$  &  $ -0.032\pm 0.003$ &  $ -0.137\pm 0.003$\\
			\hline 
		\end{tabular}
		\caption{Parameter estimation results for the fit of Euler force (see
			section~\ref{section:inertial} for details).}
		\label{tab:Eulerfit}
	\end{center}
\end{table}                

\begin{figure}
  \centering
  \includegraphics[width=0.9\linewidth]{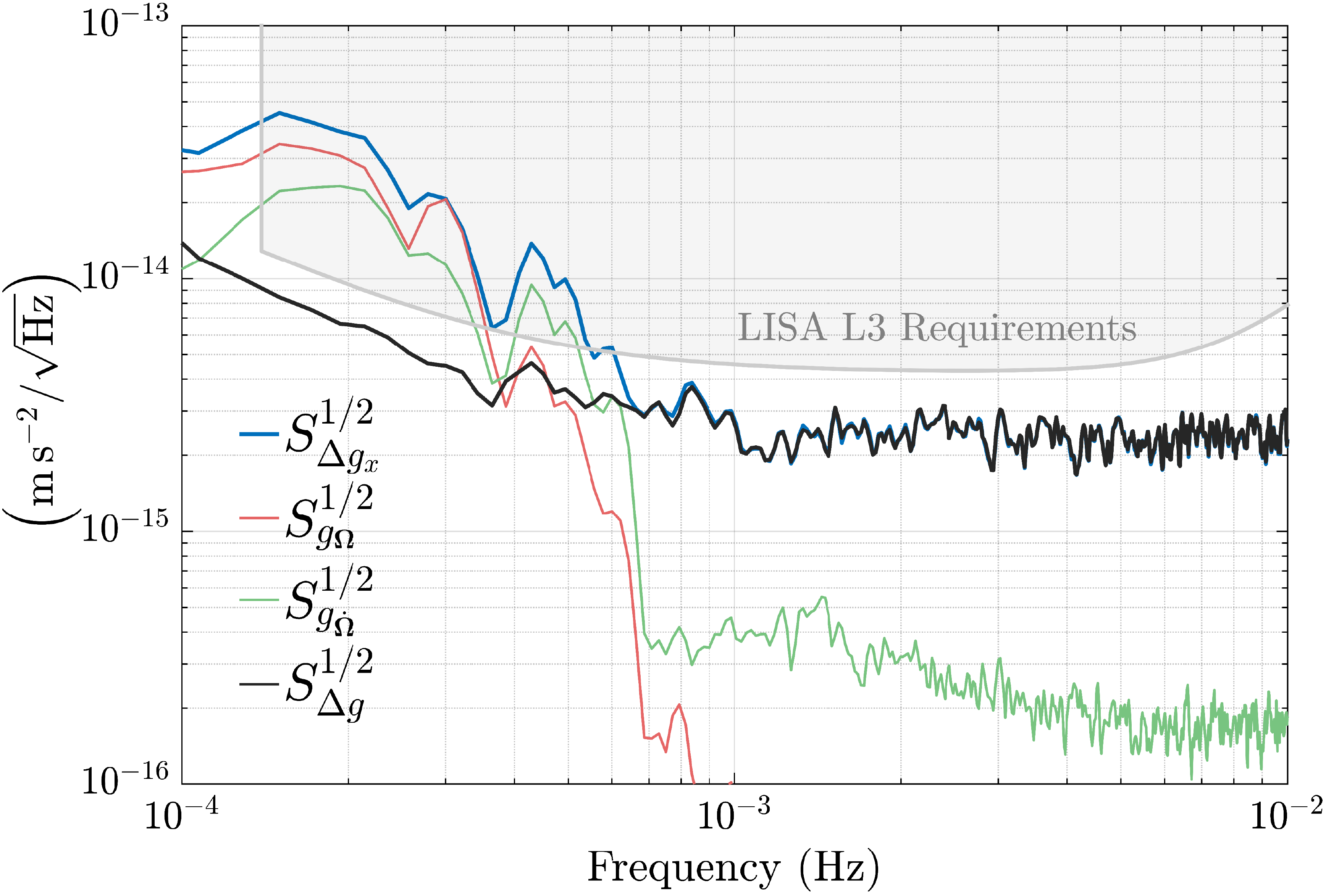}
\caption{Spectra of the breakdown of the differential acceleration noise between the two TMs at very low frequencies. Below $0.5~\mathrm{mHz}$ the effect of inertial forces on the acceleration noise is evident. Blue curve is the ASD of $\Delta g_x[t]$ calibrated as in section \ref{section:sysid} (see Fig.~\ref{fig:breakdown}), while the ASD was computed by assuming 15 data stretches overlapped by 50\%. From that curve we pass to the black one after subtracting the time series corresponding to the centrifugal term ASD (red curve) and the Euler force term ASD (green curve).}
\label{fig:inertial}
\end{figure}

To illustrate the significance of the inertial contribution subtraction, we report the effect of it on $\Delta g_x[t]$ in Figure \ref{fig:inertial} for the same data-set used in Figure~\ref{fig:breakdown}. 
For this noise data it is evident that for the low frequency part of the spectrum the subtraction is important and needed to reach the LISA requirement in that frequency band. It is also worth noting that for some other data segments the effect is less pronounced, as we would expect from the different rotational states of the spacecraft.

\begin{figure*}[!htb]
\centering
\begin{minipage}{.5\textwidth}
  \centering  
  \includegraphics[width=0.9\linewidth]{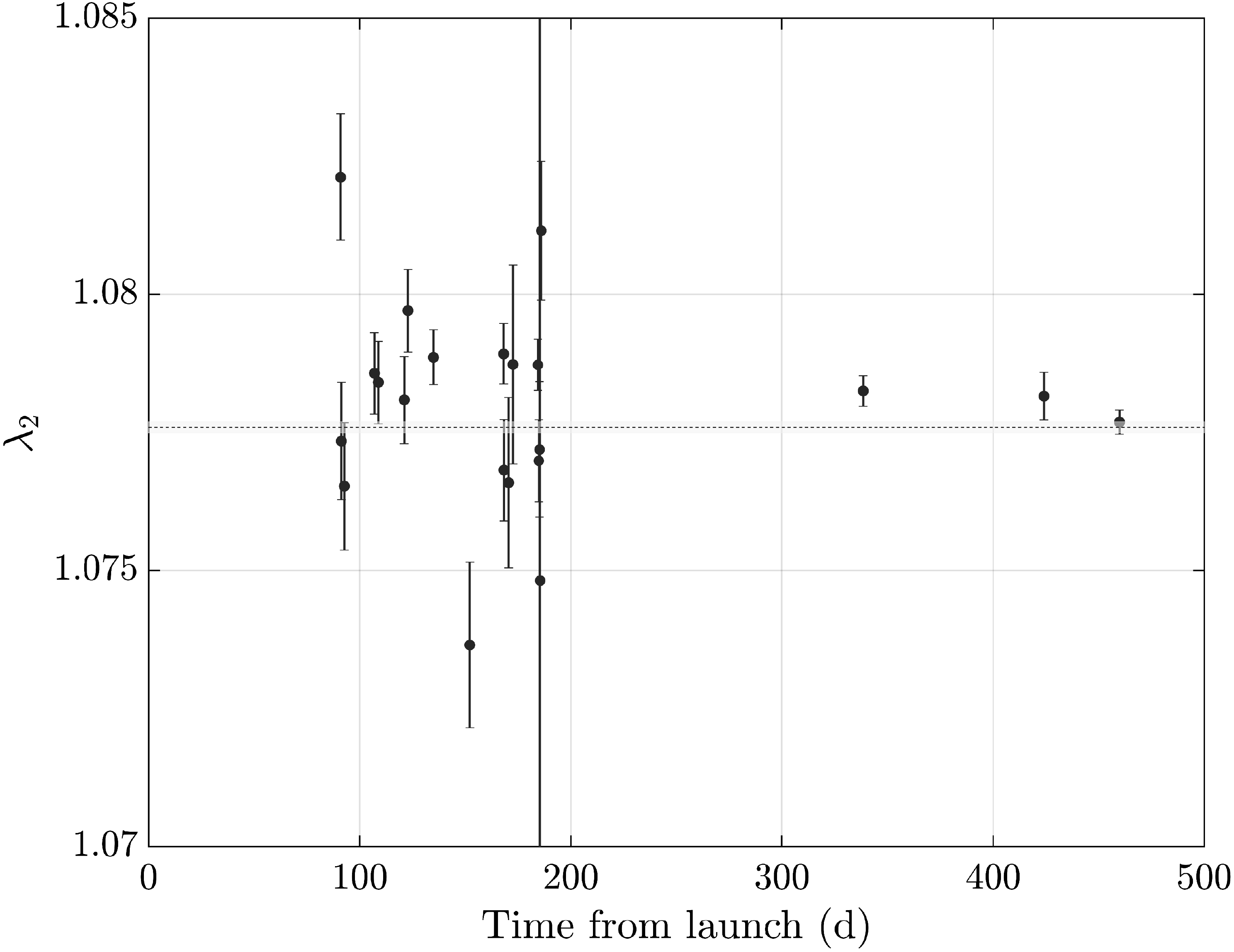}
  \caption{The $\lambda_2$ coefficient as estimated by analyzing each experiment individually (dots, with $1\sigma$ errorbars), in comparison to the value estimated assuming common value for all the experiments (dashed line). The shaded area represents the estimated error of $\lambda_2$ (see Table~\ref{tab:jointfit}  for reference).}
 \label{fig:gainevolution}
\end{minipage}%
\centering
\begin{minipage}{.5\textwidth}
  \centering  
  \includegraphics[width=0.89\linewidth]{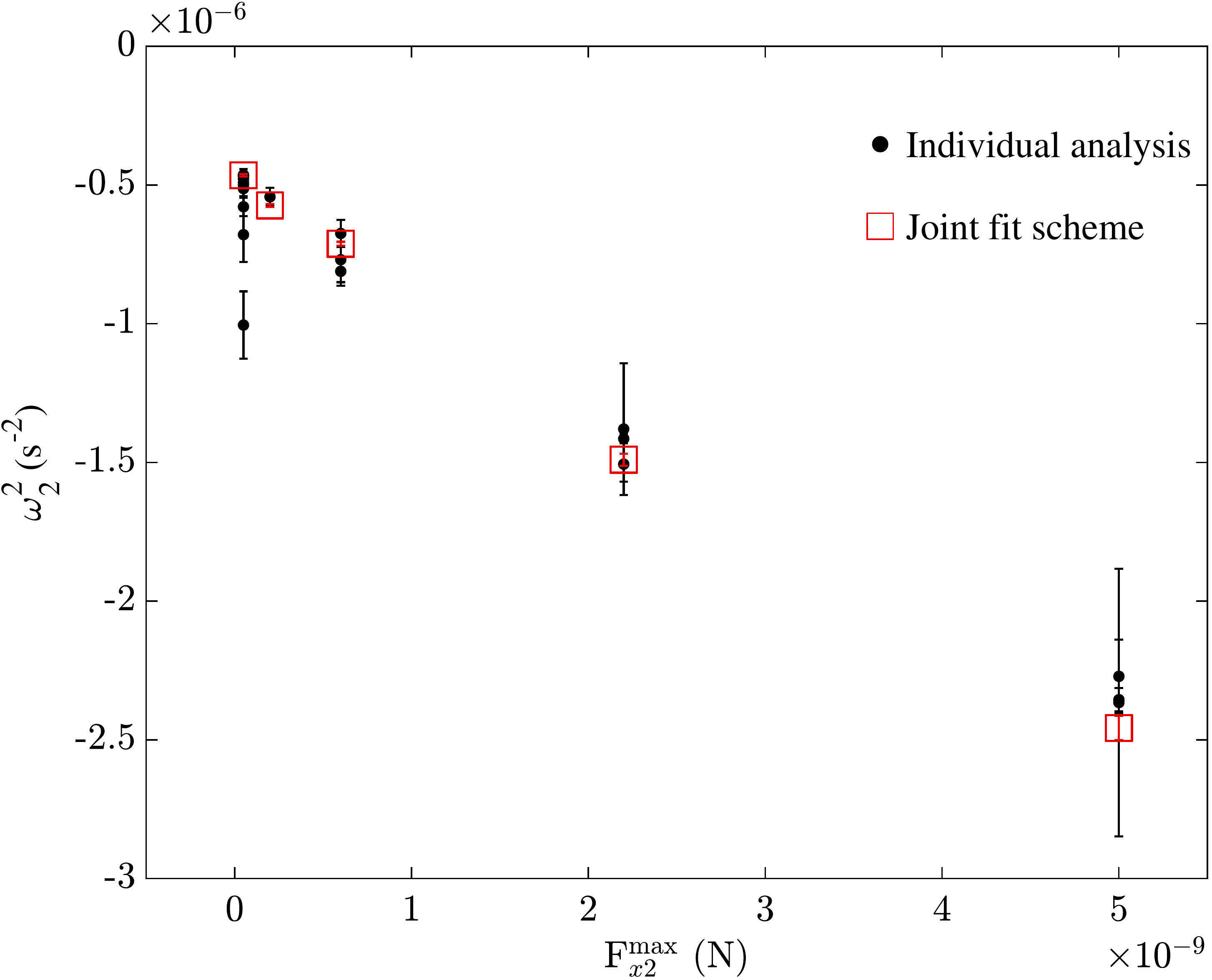}
  \caption{The measured $\omega_{2, \, \text{tot}}^2$ stiffness parameter versus the various maximum force authority configurations along the $x$-axis. 
  The black dots represent the $\omega_{2, \, \text{tot}}^2$ values estimated by analysing each experiment separately, while the red squares represent the 
  reconstructed values of $\omega_{2, \, \text{tot}}^2$, as calculated from eq.~(\ref{eq:stiff_model}) and the results from Table~\ref{tab:jointfit}.}
    \label{fig:stiffauth}
  \end{minipage}
\end{figure*}


\section{Results\label{section:results}}

The results presented in Table~\ref{tab:jointfit} show that the calibration gain coefficient of the force per unit mass applied to \tmb is found to be $\lambda_2 = 1.0776 \pm 10^{-4}$ by the joint analysis
scheme described in section~\ref{section:sysid}. This figure is in agreement with the mean value of the coefficient taken from the analysis of each 
system identification experiment individually, $\lambda_{2_\text{ind}} = 1.080 \pm 0.003$. The comparison of the two strategies of estimating the 
$\lambda_2$ is shown in Figure~\ref{fig:gainevolution} from which it is possible to deduce that the fluctuations of the calibration coefficient are always within about $0.5\%$ for the complete duration of the LPF mission. However, the reduced chi-square for a weighted average is equal to $\chi^2_{\text{red}} \simeq 90$, which is too high to claim the stationarity of $\lambda_2$. The nature of this fluctuation in $\lambda_2$ is not understood; it could be related to some true change over time, an incompleteness in our dynamical model (perhaps from a crosstalk term excited in the system identification experiments), or a residual non-linearity in the voltage actuator that manifests itself in a gain that varies slightly with the exact applied force.

The value of the gain calibration coefficient of the forces acting on TM1 was estimated to be very slightly, but significantly, smaller than $\lambda_2$, by roughly $0.3\%$ (see Table~\ref{tab:jointfit}).
This difference could be explained by the machining tolerances within the electrode housing, with gap differences of order $10~\mu\text{m}$, but also by voltage reference differences of the ADC in the feedback loop which are separate and independent for the two electrode housing. We note also that the statistical weight in extracting $\lambda_1$ is almost exclusively due to a single data segment where the \tma actuator was turned on, from 0 to $4~\text{nN}$, and not from a designed experiment dedicated to its calibration. For that reason the measurement of $\lambda_1$ can not be considered as robust as that for $\lambda_2$. 

The background residual stiffness is found to be $\omega_{1, \, 0}^2 = -(3.99\pm 0.07)\times10^{-7}~\mathrm{s}^{-2}$ 
for \tma and $\omega_{2, \, 0}^2 = -(4.19\pm 0.04)\times10^{-7}~\mathrm{s}^{-2}$ for \tmb, with their difference being compatible with zero within
 their error. The total background stiffness is estimated as the sum of the linear gravity gradient, the electrostatic stiffness from the 100 kHz capacitive sensing bias and the maximum allocated stiffness from test mass charge. Finite element calculations~\cite{0264-9381-33-23-235015} showed that the linear gravity gradients were expected to be of the order of $- 3.5\times10^{-7}$~s$^{-2}$ for \tma and $- 3.6\times10^{-7}$~s$^{-2}$ for \tmb. The electrostatic stiffness from the 100 kHz capacitive sensing bias is estimated to be $\simeq -6.4\times10^{-8}$~s$^{-2}$ and the largest charge values during these measurements is roughly $-3 \times 10^{-9} \mathrm{s}^{-2}$. The total background stiffness is  thus expected to be roughly $\simeq -4.2\times10^{-7}$~s$^{-2}$ for \tma and $\simeq -4.3\times10^{-7}$~s$^{-2}$ for \tmb, consistent with the measured values to within 10\%.

The estimates of the $\alpha_{x_j}$ and $\alpha_{\phi_j}$ are in agreement with their expected values as reported in Section~\ref{subsection:jointanalysis}, following eq.~(\ref{eq:alphax}) and~(\ref{eq:alphaphi}). In particular, for \tma, and \tmb we found that $\alpha_{x_1} = -345\pm9$~kg$^{-1}$~m$^{-1}$ 
and  $\alpha_{x_2} = -317\pm6$~kg$^{-1}$~m$^{-1}$
with a predicted value for $\alpha_{x_j}$ of -320~kg$^{-1}$~m$^{-1}$.
Similarly we find $\alpha_{\phi_1} = (-28.2\pm0.9) \times10^{3}$~kg$^{-1}$~m$^{-2}$, and 
 $\alpha_{\phi_2} = (-27.4\pm0.8) \times10^{3}$~kg$^{-1}$~m$^{-2}$, with a predicted value for $\alpha_{\phi_j}$ of
$-26.5 \times10^{3}$~kg$^{-1}$~m$^{-2}$. 

The values of the total stiffness $\omega_{j, \, \text{tot}}^2$ calculated by the theoretical model of the stiffnesses of eq.~(\ref{eq:stiff_model}) using the numerical values of Table~\ref{tab:jointfit}, is in  agreement with the estimated values of the stiffnesses by analyzing each experiment independently (see section~\ref{subsection:methodology}), proving that both methods of identifying the stiffness parameters are consistent. This is shown, as an example, for $\omega_{2}^2$ in Figure~\ref{fig:stiffauth}, where the results from 
 the two different approaches are plotted with respect to the maximum force $F_{{\rm max},2}$. This agreement confirms that the electrostatic component of the stiffnesses can be calculated from the simple model of eq.~(\ref{eq:stiff_model}), and can be used with confidence for calibration purposes in 
future space-based GW observatories.

The cross-coupling coefficient $\delta_{\text{ifo}, \, {\mathrm k}}$ describes the physical effect of signal leakage of $x_1 (t)$ to $\Delta x (t)$. This
 cross-coupling is known to be affected, to some extend, by test-mass alignment~\cite{wanner2017}, and is therefore different, as expected, for each
  of the ${\mathrm k}$ geometrical configurations of the instrument (or set-points). Estimates by numerical simulations of the optical bench including
   manufacturing tolerances, predicted an RMS value of $\simeq 6.3\times10^{-5}$ and upper and lower limits 
   to be $\simeq\pm 17\times10^{-5}$~\cite{Wanner2010thesis}. Results in Table~\ref{tab:jointfit} show values of this cross-coupling coefficient much closer to zero, indicating
    that the interferometer was indeed very well aligned during operations. Nevertheless, a deeper investigation is necessary in order to associate 
    and cross-validate the numbers of Table~\ref{tab:jointfit} with the interferometer's alignment, as well as to further investigate the 
    correlation of $\delta_{\text{ifo}, \, {\mathrm k}}$ to the cross-talk parameters originating from the SC jitter (see section~\ref{section:dynamics}). 

The low frequency spectrum of $\Delta g_x$ was found to be significantly affected by the inertial forces caused by the noisy rotation of the spacecraft. Nevertheless, we were able to directly subtract the centrifugal forces, and to fit out the Euler force contribution. This evidently improved the spectrum of $\Delta g_x$ below $0.5~\mathrm{mHz}$. However, Figure~\ref{fig:inertial} shows a residual $f^{-1}$ noise tail at those frequencies and below, which is not totally explained by the full noise model reported in~\cite{lpf_prl2} and suggests that either the model needs to be improved, or that there is an unknown low frequency noise source. Furthermore, it was not possible to totally disentangle the intrinsic degeneracy between the effect of a TM misalignment, producing a Euler force due to angular acceleration, and that due to electrostatic actuation cross-talk. However, there is evidence that during periods of strong roll of the spacecraft the effect of the subtraction is more pronounced, suggesting that the major part of the effect is really due to the Euler force contribution.


\section{Discussion\label{section:conclusions}}

We have performed a set of calibration experiments on-board the LPF satellite. The aim of these experiments was to determine the 
dynamics of the three-body system and to investigate the stationarity and performance of the hardware. 
Two different data analysis strategies were adopted. First, during mission operations we performed $\chi^2$ fits on each of the experiments, which were reliable and computationally cheap. Secondly, by assuming a generalized model on the stiffnesses depending on the actuation authority, we were able to adopt a joint fitting scheme on the complete set of all calibration experiments performed over the full duration of the mission.
Both approaches yielded consistent results, and the dynamical 
parameters were estimated to be in agreement with the expected values.   We were also able to determine the contribution due to inertial forces and
 to subtract them from the calibrated $\Delta g_x$ for the mission duration. This has improved significantly the low frequency spectrum of the acceleration noise.
The full calibration $\Delta g$ procedure has proved essential for LPF to reach the exceptional noise levels shown in~\cite{lpf_prl}. Moreover, it was the baseline 
starting point for many other investigations performed on board LPF. 

Space-borne GW detectors such as LISA will inherit the hardware technology directly from the LPF mission, and the
same type of calibration experiments could, in principle, be applied to the three spacecrafts configuration as well.
Indeed, for the case of LISA, any differential force between two free falling TMs that
cannot be associated  to an actual stray force acting on the TMs could be entangled with GW signals,
and therefore it will be necessary to estimate and subtract it. However in LISA, unlike in LPF, all the TMs will be drag-free along their $x$-axis, so 
there will be no intentional electrostatic force applied on the TMs along the sensitive $x$-axis. Thus for
a fully functioning LISA constellation operating in science mode, a gain actuator calibration along $x$ is not needed. 

However, in addition to normal operation of LISA, there are potential fall-back cases, such as the failure of one optical link, where a TM could be electrostatically suspended along $x$, and therefore calibration of the applied $x$ force would be necessary in order to save the scientific outcome of the observatory. Calibrating the electrostatic force along the $x$-axis in LISA has one major difference with respect to the LPF case. Indeed, the induced force signals will propagate through the long arms of the constellation, thus requiring the full LISA interferometric read out system of time delay interferometry (TDI)~\cite{Tinto2005} to be employed. However the principle of the measurements is exactly the same, that of stimulating the system via injecting large modulation signals. Even if there are no electrostatic forces on the TMs along the sensitive $x$-axes, all the TMs will be electrostatically suspended on the remaining degrees of freedom in order to control each spacecraft to follow the two TMs it encloses. For those degrees of freedom an actuator calibration would be necessary, and the calibration experiments could, in-principle, follow the same philosophy as the ones performed for LPF (described in Sec.~\ref{section:sysid}). Moreover, there will still be force gradients present in the environment of the satellites in the same way as they were present in LPF. In order to characterize these gradients, an injection along $x$ into the drag-free control loop, or an out of the loop force stimulus on the satellites would be necessary to estimate the stiffness, so a calibration experiment along the sensitive axes on the same line of those described in this paper will be needed as well. 
%
 
Regarding the possible appearance of inertial forces acting on the TMs, a first analysis shows that the incidence of those types of contribution in LISA should be mitigated 
by the LISA attitude control, which operates with respect to the far away spacecraft laser source. The precision of the angular measurement should be of the order of $\mathrm{nrad}/\sqrt{\mathrm{Hz}}$, while for the AST was of the order of $\mathrm{mrad}/\sqrt{\mathrm{Hz}}$~\cite{lpf_prl}. However, further investigation on the possible impact of inertial forces on LISA is also ongoing.


\begin{acknowledgments}
This work has been made possible by the LISA Pathfinder mission, which is part of the
space-science programme of the European Space Agency.

The French contribution has been supported by the CNES (Accord Specific de projet
CNES 1316634/CNRS 103747), the CNRS, the Observatoire de Paris and the University
Paris-Diderot. E.~Plagnol and H.~Inchausp\'{e} would also like to acknowledge the
financial support of the UnivEarthS Labex program at Sorbonne Paris Cit\'{e}
(ANR-10-LABX-0023 and ANR-11-IDEX-0005-02).

The Albert-Einstein-Institut acknowledges the support of the German Space Agency,
DLR. The work is supported by the Federal Ministry for Economic Affairs and Energy
based on a resolution of the German Bundestag (FKZ 50OQ0501 and FKZ 50OQ1601). 

The Italian contribution has been supported  by Agenzia Spaziale Italiana and Istituto
Nazionale di Fisica Nucleare.

The Spanish contribution has been supported by contracts AYA2010-15709 (MICINN),
ESP2013-47637-P, and ESP2015-67234-P (MINECO). M.~Nofrarias acknowledges support from 
Fundacion General CSIC (Programa ComFuturo). F.~Rivas acknowledges an FPI contract
(MINECO).

The Swiss contribution acknowledges the support of the Swiss Space Office (SSO)
via the PRODEX Programme of ESA. L.~Ferraioli acknowledges the support of the Swiss National Science Foundation.
N. Meshskar acknowledges the support of the ETH Zurich (ETH-05 16-2).

The UK groups wish to acknowledge support from the United Kingdom Space Agency
(UKSA), the University of Glasgow, the University of Birmingham, Imperial College,
and the Scottish Universities Physics Alliance (SUPA).

J.\,I.~Thorpe and J.~Slutsky acknowledge the support of the US National Aeronautics
and Space Administration (NASA).

N. Korsakova would like to acknowledge the support of the Newton International Fellowship from the Royal Society.
\end{acknowledgments}


\smallskip


\medskip
\section*{References}

\begin{thebibliography}{29}
\providecommand{\natexlab}[1]{#1}
\providecommand{\url}[1]{\texttt{#1}}
\expandafter\ifx\csname urlstyle\endcsname\relax
  \providecommand{\doi}[1]{doi: #1}\else
  \providecommand{\doi}{doi: \begingroup \urlstyle{rm}\Url}\fi

\bibitem[McNamara et~al.(2008)McNamara, Vitale, Danzmann, and on~behalf of~the
  LISA Pathfinder Science Working~Team]{0264-9381-25-11-114034}
P~McNamara, S~Vitale, K~Danzmann, and on~behalf of~the LISA Pathfinder Science
  Working~Team.
\newblock LISA Pathfinder.
\newblock \emph{Classical and Quantum Gravity}, 25\penalty0 (11):\penalty0
  114034, 2008.
\newblock URL \url{http://stacks.iop.org/0264-9381/25/i=11/a=114034}.

\bibitem[Antonucci et~al.(2011)Antonucci, Armano, Audley, Auger, Benedetti,
  Binetruy, Boatella, Bogenstahl, Bortoluzzi, Bosetti, Brandt, Caleno,
  Cavalleri, Cesa, Chmeissani, Ciani, Conchillo, Congedo, Cristofolini, Cruise,
  Danzmann, Marchi, Diaz-Aguilo, Diepholz, Dixon, Dolesi, Dunbar, Fauste,
  Ferraioli, Fertin, Fichter, Fitzsimons, Freschi, Marin, Marirrodriga, Gerndt,
  Gesa, Giardini, Gibert, Grimani, Grynagier, Guillaume, Guzmán, Harrison,
  Heinzel, Hewitson, Hollington, Hough, Hoyland, Hueller, Huesler, Jeannin,
  Jennrich, Jetzer, Johlander, Killow, Llamas, Lloro, Lobo, Maarschalkerweerd,
  Madden, Mance, Mateos, McNamara, Mendes, Mitchell, Monsky, Nicolini,
  Nicolodi, Nofrarias, Pedersen, Perreur-Lloyd, Perreca, Plagnol, Prat, Racca,
  Rais, Ramos-Castro, Reiche, Perez, Robertson, Rozemeijer, Sanjuan,
  Schleicher, Schulte, Shaul, Stagnaro, Strandmoe, Steier, Sumner, Taylor,
  Texier, Trenkel, Tombolato, Vitale, Wanner, Ward, Waschke, Wass, Weber, and
  Zweifel]{lpf1}
F~Antonucci, M~Armano, H~Audley, G~Auger, M~Benedetti, P~Binetruy, C~Boatella,
  J~Bogenstahl, D~Bortoluzzi, P~Bosetti, N~Brandt, M~Caleno, A~Cavalleri,
  M~Cesa, M~Chmeissani, G~Ciani, A~Conchillo, G~Congedo, I~Cristofolini,
  M~Cruise, K~Danzmann, F~De Marchi, M~Diaz-Aguilo, I~Diepholz, G~Dixon,
  R~Dolesi, N~Dunbar, J~Fauste, L~Ferraioli, D~Fertin, W~Fichter, E~Fitzsimons,
  M~Freschi, A~García Marin, C~García Marirrodriga, R~Gerndt, L~Gesa,
  D~Giardini, F~Gibert, C~Grimani, A~Grynagier, B~Guillaume, F~Guzmán,
  I~Harrison, G~Heinzel, M~Hewitson, D~Hollington, J~Hough, D~Hoyland,
  M~Hueller, J~Huesler, O~Jeannin, O~Jennrich, P~Jetzer, B~Johlander, C~Killow,
  X~Llamas, I~Lloro, A~Lobo, R~Maarschalkerweerd, S~Madden, D~Mance, I~Mateos,
  P~W McNamara, J~Mendes, E~Mitchell, A~Monsky, D~Nicolini, D~Nicolodi,
  M~Nofrarias, F~Pedersen, M~Perreur-Lloyd, A~Perreca, E~Plagnol, P~Prat, G~D
  Racca, B~Rais, J~Ramos-Castro, J~Reiche, J~A~Romera Perez, D~Robertson,
  H~Rozemeijer, J~Sanjuan, A~Schleicher, M~Schulte, D~Shaul, L~Stagnaro,
  S~Strandmoe, F~Steier, T~J Sumner, A~Taylor, D~Texier, C~Trenkel,
  D~Tombolato, S~Vitale, G~Wanner, H~Ward, S~Waschke, P~Wass, W~J Weber, and
  P~Zweifel.
\newblock From laboratory experiments to LISA Pathfinder: achieving LISA
  geodesic motion.
\newblock \emph{Classical and Quantum Gravity}, 28\penalty0 (9):\penalty0
  094002, 2011.
\newblock URL \url{http://stacks.iop.org/0264-9381/28/i=9/a=094002}.

\bibitem[{Amaro-Seoane} et~al.(2017){Amaro-Seoane}, {Audley}, {Babak}, {Baker},
  {Barausse}, {Bender}, {Berti}, {Binetruy}, {Born}, {Bortoluzzi}, {Camp},
  {Caprini}, {Cardoso}, {Colpi}, {Conklin}, {Cornish}, {Cutler}, {Danzmann},
  {Dolesi}, {Ferraioli}, {Ferroni}, {Fitzsimons}, {Gair}, {Gesa Bote},
  {Giardini}, {Gibert}, {Grimani}, {Halloin}, {Heinzel}, {Hertog}, {Hewitson},
  {Holley-Bockelmann}, {Hollington}, {Hueller}, {Inchauspe}, {Jetzer},
  {Karnesis}, {Killow}, {Klein}, {Klipstein}, {Korsakova}, {Larson}, {Livas},
  {Lloro}, {Man}, {Mance}, {Martino}, {Mateos}, {McKenzie}, {McWilliams},
  {Miller}, {Mueller}, {Nardini}, {Nelemans}, {Nofrarias}, {Petiteau},
  {Pivato}, {Plagnol}, {Porter}, {Reiche}, {Robertson}, {Robertson}, {Rossi},
  {Russano}, {Schutz}, {Sesana}, {Shoemaker}, {Slutsky}, {Sopuerta}, {Sumner},
  {Tamanini}, {Thorpe}, {Troebs}, {Vallisneri}, {Vecchio}, {Vetrugno},
  {Vitale}, {Volonteri}, {Wanner}, {Ward}, {Wass}, {Weber}, {Ziemer}, and
  {Zweifel}]{2017arXiv170200786A}
P.~{Amaro-Seoane}, H.~{Audley}, S.~{Babak}, J.~{Baker}, E.~{Barausse},
  P.~{Bender}, E.~{Berti}, P.~{Binetruy}, M.~{Born}, D.~{Bortoluzzi},
  J.~{Camp}, C.~{Caprini}, V.~{Cardoso}, M.~{Colpi}, J.~{Conklin},
  N.~{Cornish}, C.~{Cutler}, K.~{Danzmann}, R.~{Dolesi}, L.~{Ferraioli},
  V.~{Ferroni}, E.~{Fitzsimons}, J.~{Gair}, L.~{Gesa Bote}, D.~{Giardini},
  F.~{Gibert}, C.~{Grimani}, H.~{Halloin}, G.~{Heinzel}, T.~{Hertog},
  M.~{Hewitson}, K.~{Holley-Bockelmann}, D.~{Hollington}, M.~{Hueller},
  H.~{Inchauspe}, P.~{Jetzer}, N.~{Karnesis}, C.~{Killow}, A.~{Klein},
  B.~{Klipstein}, N.~{Korsakova}, S.~L {Larson}, J.~{Livas}, I.~{Lloro},
  N.~{Man}, D.~{Mance}, J.~{Martino}, I.~{Mateos}, K.~{McKenzie}, S.~T
  {McWilliams}, C.~{Miller}, G.~{Mueller}, G.~{Nardini}, G.~{Nelemans},
  M.~{Nofrarias}, A.~{Petiteau}, P.~{Pivato}, E.~{Plagnol}, E.~{Porter},
  J.~{Reiche}, D.~{Robertson}, N.~{Robertson}, E.~{Rossi}, G.~{Russano},
  B.~{Schutz}, A.~{Sesana}, D.~{Shoemaker}, J.~{Slutsky}, C.~F. {Sopuerta},
  T.~{Sumner}, N.~{Tamanini}, I.~{Thorpe}, M.~{Troebs}, M.~{Vallisneri},
  A.~{Vecchio}, D.~{Vetrugno}, S.~{Vitale}, M.~{Volonteri}, G.~{Wanner},
  H.~{Ward}, P.~{Wass}, W.~{Weber}, J.~{Ziemer}, and P.~{Zweifel}.
\newblock {Laser Interferometer Space Antenna}.
\newblock \emph{ArXiv e-prints}, February 2017.
\newblock \url{https://arxiv.org/abs/1702.00786}

\bibitem[Armano et~al.(2016{\natexlab{a}})Armano, Audley, Auger, Baird, Bassan,
  Binetruy, Born, Bortoluzzi, Brandt, Caleno, Carbone, Cavalleri, Cesarini,
  Ciani, Congedo, Cruise, Danzmann, de~Deus~Silva, De~Rosa, Diaz-Aguil\'o,
  Di~Fiore, Diepholz, Dixon, Dolesi, Dunbar, Ferraioli, Ferroni, Fichter,
  Fitzsimons, Flatscher, Freschi, Garc\'{\i}a~Mar\'{\i}n,
  Garc\'{\i}a~Marirrodriga, Gerndt, Gesa, Gibert, Giardini, Giusteri, Guzm\'an,
  Grado, Grimani, Grynagier, Grzymisch, Harrison, Heinzel, Hewitson,
  Hollington, Hoyland, Hueller, Inchausp\'e, Jennrich, Jetzer, Johann,
  Johlander, Karnesis, Kaune, Korsakova, Killow, Lobo, Lloro, Liu,
  L\'opez-Zaragoza, Maarschalkerweerd, Mance, Mart\'{\i}n, Martin-Polo,
  Martino, Martin-Porqueras, Madden, Mateos, McNamara, Mendes, Mendes, Monsky,
  Nicolodi, Nofrarias, Paczkowski, Perreur-Lloyd, Petiteau, Pivato, Plagnol,
  Prat, Ragnit, Ra\"{\i}s, Ramos-Castro, Reiche, Robertson, Rozemeijer, Rivas,
  Russano, Sanju\'an, Sarra, Schleicher, Shaul, Slutsky, Sopuerta, Stanga,
  Steier, Sumner, Texier, Thorpe, Trenkel, Tr\"obs, Tu, Vetrugno, Vitale, Wand,
  Wanner, Ward, Warren, Wass, Wealthy, Weber, Wissel, Wittchen, Zambotti,
  Zanoni, Ziegler, and Zweifel]{lpf_prl}
M.~Armano, H.~Audley, G.~Auger, J.~T. Baird, M.~Bassan, P.~Binetruy, M.~Born,
  D.~Bortoluzzi, N.~Brandt, M.~Caleno, L.~Carbone, A.~Cavalleri, A.~Cesarini,
  G.~Ciani, G.~Congedo, A.~M. Cruise, K.~Danzmann, M.~de~Deus~Silva,
  R.~De~Rosa, M.~Diaz-Aguil\'o, L.~Di~Fiore, I.~Diepholz, G.~Dixon, R.~Dolesi,
  N.~Dunbar, L.~Ferraioli, V.~Ferroni, W.~Fichter, E.~D. Fitzsimons,
  R.~Flatscher, M.~Freschi, A.~F. Garc\'{\i}a~Mar\'{\i}n,
  C.~Garc\'{\i}a~Marirrodriga, R.~Gerndt, L.~Gesa, F.~Gibert, D.~Giardini,
  R.~Giusteri, F.~Guzm\'an, A.~Grado, C.~Grimani, A.~Grynagier, J.~Grzymisch,
  I.~Harrison, G.~Heinzel, M.~Hewitson, D.~Hollington, D.~Hoyland, M.~Hueller,
  H.~Inchausp\'e, O.~Jennrich, P.~Jetzer, U.~Johann, B.~Johlander, N.~Karnesis,
  B.~Kaune, N.~Korsakova, C.~J. Killow, J.~A. Lobo, I.~Lloro, L.~Liu, J.~P.
  L\'opez-Zaragoza, R.~Maarschalkerweerd, D.~Mance, V.~Mart\'{\i}n,
  L.~Martin-Polo, J.~Martino, F.~Martin-Porqueras, S.~Madden, I.~Mateos, P.~W.
  McNamara, J.~Mendes, L.~Mendes, A.~Monsky, D.~Nicolodi, M.~Nofrarias,
  S.~Paczkowski, M.~Perreur-Lloyd, A.~Petiteau, P.~Pivato, E.~Plagnol, P.~Prat,
  U.~Ragnit, B.~Ra\"{\i}s, J.~Ramos-Castro, J.~Reiche, D.~I. Robertson,
  H.~Rozemeijer, F.~Rivas, G.~Russano, J.~Sanju\'an, P.~Sarra, A.~Schleicher,
  D.~Shaul, J.~Slutsky, C.~F. Sopuerta, R.~Stanga, F.~Steier, T.~Sumner,
  D.~Texier, J.~I. Thorpe, C.~Trenkel, M.~Tr\"obs, H.~B. Tu, D.~Vetrugno,
  S.~Vitale, V.~Wand, G.~Wanner, H.~Ward, C.~Warren, P.~J. Wass, D.~Wealthy,
  W.~J. Weber, L.~Wissel, A.~Wittchen, A.~Zambotti, C.~Zanoni, T.~Ziegler, and
  P.~Zweifel.
\newblock Sub-femto-$g$ free fall for space-based gravitational wave
  observatories: LISA Pathfinder results.
\newblock \emph{Phys. Rev. Lett.}, 116:\penalty0 231101, Jun
  2016{\natexlab{a}}.
\newblock \doi{10.1103/PhysRevLett.116.231101}.
\newblock URL \url{http://link.aps.org/doi/10.1103/PhysRevLett.116.231101}.

\bibitem[Armano et~al.(2018)Armano, Audley, Baird, Binetruy, Born, Bortoluzzi,
  Castelli, Cavalleri, Cesarini, Cruise, Danzmann, de~Deus~Silva, Diepholz,
  Dixon, Dolesi, Ferraioli, Ferroni, Fitzsimons, Freschi, Gesa, Gibert,
  Giardini, Giusteri, Grimani, Grzymisch, Harrison, Heinzel, Hewitson,
  Hollington, Hoyland, Hueller, Inchausp\'e, Jennrich, Jetzer, Karnesis, Kaune,
  Korsakova, Killow, Lobo, Lloro, Liu, L\'opez-Zaragoza, Maarschalkerweerd,
  Mance, Meshksar, Mart\'{\i}n, Martin-Polo, Martino, Martin-Porqueras, Mateos,
  McNamara, Mendes, Mendes, Nofrarias, Paczkowski, Perreur-Lloyd, Petiteau,
  Pivato, Plagnol, Ramos-Castro, Reiche, Robertson, Rivas, Russano, Slutsky,
  Sopuerta, Sumner, Texier, Thorpe, Vetrugno, Vitale, Wanner, Ward, Wass,
  Weber, Wissel, Wittchen, and Zweifel]{lpf_prl2}
M.~Armano, H.~Audley, J.~Baird, P.~Binetruy, M.~Born, D.~Bortoluzzi,
  E.~Castelli, A.~Cavalleri, A.~Cesarini, A.~M. Cruise, K.~Danzmann,
  M.~de~Deus~Silva, I.~Diepholz, G.~Dixon, R.~Dolesi, L.~Ferraioli, V.~Ferroni,
  E.~D. Fitzsimons, M.~Freschi, L.~Gesa, F.~Gibert, D.~Giardini, R.~Giusteri,
  C.~Grimani, J.~Grzymisch, I.~Harrison, G.~Heinzel, M.~Hewitson,
  D.~Hollington, D.~Hoyland, M.~Hueller, H.~Inchausp\'e, O.~Jennrich,
  P.~Jetzer, N.~Karnesis, B.~Kaune, N.~Korsakova, C.~J. Killow, J.~A. Lobo,
  I.~Lloro, L.~Liu, J.~P. L\'opez-Zaragoza, R.~Maarschalkerweerd, D.~Mance,
  N.~Meshksar, V.~Mart\'{\i}n, L.~Martin-Polo, J.~Martino, F.~Martin-Porqueras,
  I.~Mateos, P.~W. McNamara, J.~Mendes, L.~Mendes, M.~Nofrarias, S.~Paczkowski,
  M.~Perreur-Lloyd, A.~Petiteau, P.~Pivato, E.~Plagnol, J.~Ramos-Castro,
  J.~Reiche, D.~I. Robertson, F.~Rivas, G.~Russano, J.~Slutsky, C.~F. Sopuerta,
  T.~Sumner, D.~Texier, J.~I. Thorpe, D.~Vetrugno, S.~Vitale, G.~Wanner,
  H.~Ward, P.~J. Wass, W.~J. Weber, L.~Wissel, A.~Wittchen, and P.~Zweifel.
\newblock Beyond the required LISA free-fall performance: New LISA Pathfinder
  results down to $20\text{ }\text{ }\ensuremath{\mu}\mathrm{Hz}$.
\newblock \emph{Phys. Rev. Lett.}, 120:\penalty0 061101, Feb 2018.
\newblock \doi{10.1103/PhysRevLett.120.061101}.
\newblock URL \url{https://link.aps.org/doi/10.1103/PhysRevLett.120.061101}.

\bibitem[Hewitson et~al.(2009)Hewitson, Armano, Benedetti, Bogenstahl,
  Bortoluzzi, Bosetti, Brandt, Cavalleri, Ciani, Cristofolini, Cruise,
  Danzmann, Diepholz, Dolesi, Fauste, Ferraioli, Fertin, Fichter, García,
  García, Grynagier, Guzmán, Fitzsimons, Heinzel, Hollington, Hough, Hueller,
  Hoyland, Jennrich, Johlander, Killow, Lobo, Mance, Mateos, McNamara, Monsky,
  Nicolini, Nicolodi, Nofrarias, Perreur-Lloyd, Plagnol, Racca, Ramos-Castro,
  Robertson, Sanjuan, Schulte, Shaul, Smit, Stagnaro, Steier, Sumner, Tateo,
  Tombolato, Vischer, Vitale, Wanner, Ward, Waschke, Wand, Wass, Weber,
  Ziegler, and Zweifel]{ltpda}
M~Hewitson, M~Armano, M~Benedetti, J~Bogenstahl, D~Bortoluzzi, P~Bosetti,
  N~Brandt, A~Cavalleri, G~Ciani, I~Cristofolini, M~Cruise, K~Danzmann,
  I~Diepholz, R~Dolesi, J~Fauste, L~Ferraioli, D~Fertin, W~Fichter, A~García,
  C~García, A~Grynagier, F~Guzmán, E~Fitzsimons, G~Heinzel, D~Hollington,
  J~Hough, M~Hueller, D~Hoyland, O~Jennrich, B~Johlander, C~Killow, A~Lobo,
  D~Mance, I~Mateos, P~W McNamara, A~Monsky, D~Nicolini, D~Nicolodi,
  M~Nofrarias, M~Perreur-Lloyd, E~Plagnol, G~D Racca, J~Ramos-Castro,
  D~Robertson, J~Sanjuan, M~O Schulte, D~N~A Shaul, M~Smit, L~Stagnaro,
  F~Steier, T~J Sumner, N~Tateo, D~Tombolato, G~Vischer, S~Vitale, G~Wanner,
  H~Ward, S~Waschke, V~Wand, P~Wass, W~J Weber, T~Ziegler, and P~Zweifel.
\newblock Data analysis for the LISA technology package.
\newblock \emph{Classical and Quantum Gravity}, 26\penalty0 (9):\penalty0
  094003, 2009.
\newblock URL \url{http://stacks.iop.org/0264-9381/26/i=9/a=094003}.

\bibitem[Gerndt and the~entire LTP~Team(2009)]{1742-6596-154-1-012007}
R~Gerndt and the~entire LTP~Team.
\newblock LTP – LISA technology package: Development challenges of a
  spaceborne fundamental physics experiment.
\newblock \emph{Journal of Physics: Conference Series}, 154\penalty0
  (1):\penalty0 012007, 2009.
\newblock URL \url{http://stacks.iop.org/1742-6596/154/i=1/a=012007}.

\bibitem[Anza et~al.(2005)Anza, Armano, Balaguer, Benedetti, Boatella, Bosetti,
  Bortoluzzi, Brandt, Braxmaier, Caldwell, Carbone, Cavalleri, Ciccolella,
  Cristofolini, Cruise, Lio, Danzmann, Desiderio, Dolesi, Dunbar, Fichter,
  Garcia, Garcia-Berro, Marin, Gerndt, Gianolio, Giardini, Gruenagel,
  Hammesfahr, Heinzel, Hough, Hoyland, Hueller, Jennrich, Johann, Kemble,
  Killow, Kolbe, Landgraf, Lobo, Lorizzo, Mance, Middleton, Nappo, Nofrarias,
  Racca, Ramos, Robertson, Sallusti, Sandford, Sanjuan, Sarra, Selig, Shaul,
  Smart, Smit, Stagnaro, Sumner, Tirabassi, Tobin, Vitale, Wand, Ward, Weber,
  and Zweifel]{0264-9381-22-10-001}
S~Anza, M~Armano, E~Balaguer, M~Benedetti, C~Boatella, P~Bosetti, D~Bortoluzzi,
  N~Brandt, C~Braxmaier, M~Caldwell, L~Carbone, A~Cavalleri, A~Ciccolella,
  I~Cristofolini, M~Cruise, M~Da Lio, K~Danzmann, D~Desiderio, R~Dolesi,
  N~Dunbar, W~Fichter, C~Garcia, E~Garcia-Berro, A~F~Garcia Marin, R~Gerndt,
  A~Gianolio, D~Giardini, R~Gruenagel, A~Hammesfahr, G~Heinzel, J~Hough,
  D~Hoyland, M~Hueller, O~Jennrich, U~Johann, S~Kemble, C~Killow, D~Kolbe,
  M~Landgraf, A~Lobo, V~Lorizzo, D~Mance, K~Middleton, F~Nappo, M~Nofrarias,
  G~Racca, J~Ramos, D~Robertson, M~Sallusti, M~Sandford, J~Sanjuan, P~Sarra,
  A~Selig, D~Shaul, D~Smart, M~Smit, L~Stagnaro, T~Sumner, C~Tirabassi,
  S~Tobin, S~Vitale, V~Wand, H~Ward, W~J Weber, and P~Zweifel.
\newblock The LTP experiment on the LISA Pathfinder mission.
\newblock \emph{Classical and Quantum Gravity}, 22\penalty0 (10):\penalty0
  S125, 2005.
\newblock URL \url{http://stacks.iop.org/0264-9381/22/i=10/a=001}.

\bibitem[Robertson et~al.(2013)Robertson, Fitzsimons, Killow, Perreur-Lloyd,
  Ward, Bryant, Cruise, Dixon, Hoyland, Smith, and Bogenstahl]{oms4}
D~I Robertson, E~D Fitzsimons, C~J Killow, M~Perreur-Lloyd, H~Ward, J~Bryant,
  A~M Cruise, G~Dixon, D~Hoyland, D~Smith, and J~Bogenstahl.
\newblock Construction and testing of the optical bench for LISA Pathfinder.
\newblock \emph{Classical and Quantum Gravity}, 30\penalty0 (8):\penalty0
  085006, 2013.
\newblock URL \url{http://stacks.iop.org/0264-9381/30/i=8/a=085006}.

\bibitem[Heinzel et~al.(2003)Heinzel, Braxmaier, Schilling, Rüdiger,
  Robertson, te~Plate, Wand, Arai, Johann, and Danzmann]{oms3}
G~Heinzel, C~Braxmaier, R~Schilling, A~Rüdiger, D~Robertson, M~te~Plate,
  V~Wand, K~Arai, U~Johann, and K~Danzmann.
\newblock Interferometry for the LISA technology package (LTP) aboard smart-2.
\newblock \emph{Classical and Quantum Gravity}, 20\penalty0 (10):\penalty0
  S153, 2003.
\newblock URL \url{http://stacks.iop.org/0264-9381/20/i=10/a=318}.

\bibitem[Heinzel et~al.(2004)Heinzel, Wand, García, Jennrich, Braxmaier,
  Robertson, Middleton, Hoyland, Rüdiger, Schilling, Johann, and
  Danzmann]{oms2}
G~Heinzel, V~Wand, A~García, O~Jennrich, C~Braxmaier, D~Robertson,
  K~Middleton, D~Hoyland, A~Rüdiger, R~Schilling, U~Johann, and K~Danzmann.
\newblock The LTP interferometer and phasemeter.
\newblock \emph{Classical and Quantum Gravity}, 21\penalty0 (5):\penalty0 S581,
  2004.
\newblock URL \url{http://stacks.iop.org/0264-9381/21/i=5/a=029}.

\bibitem[Audley et~al.(2011)Audley, Danzmann, Marín, Heinzel, Monsky,
  Nofrarias, Steier, Gerardi, Gerndt, Hechenblaikner, Johann, Luetzow-Wentzky,
  Wand, Antonucci, Armano, Auger, Benedetti, Binetruy, Boatella, Bogenstahl,
  Bortoluzzi, Bosetti, Caleno, Cavalleri, Cesa, Chmeissani, Ciani, Conchillo,
  Congedo, Cristofolini, Cruise, Marchi, Diaz-Aguilo, Diepholz, Dixon, Dolesi,
  Fauste, Ferraioli, Fertin, Fichter, Fitzsimons, Freschi, Marirrodriga, Gesa,
  Gibert, Giardini, Grimani, Grynagier, Guillaume, Guzmán, Harrison, Hewitson,
  Hollington, Hough, Hoyland, Hueller, Huesler, Jeannin, Jennrich, Jetzer,
  Johlander, Killow, Llamas, Lloro, Lobo, Maarschalkerweerd, Madden, Mance,
  Mateos, McNamara, Mendes, Mitchell, Nicolini, Nicolodi, Pedersen,
  Perreur-Lloyd, Perreca, Plagnol, Prat, Racca, Rais, Ramos-Castro, Reiche,
  Perez, Robertson, Rozemeijer, Sanjuan, Schulte, Shaul, Stagnaro, Strandmoe,
  Sumner, Taylor, Texier, Trenkel, Tombolato, Vitale, Wanner, Ward, Waschke,
  Wass, Weber, and Zweifel]{oms1}
H~Audley, K~Danzmann, A~García Marín, G~Heinzel, A~Monsky, M~Nofrarias,
  F~Steier, D~Gerardi, R~Gerndt, G~Hechenblaikner, U~Johann, P~Luetzow-Wentzky,
  V~Wand, F~Antonucci, M~Armano, G~Auger, M~Benedetti, P~Binetruy, C~Boatella,
  J~Bogenstahl, D~Bortoluzzi, P~Bosetti, M~Caleno, A~Cavalleri, M~Cesa,
  M~Chmeissani, G~Ciani, A~Conchillo, G~Congedo, I~Cristofolini, M~Cruise, F~De
  Marchi, M~Diaz-Aguilo, I~Diepholz, G~Dixon, R~Dolesi, J~Fauste, L~Ferraioli,
  D~Fertin, W~Fichter, E~Fitzsimons, M~Freschi, C~García Marirrodriga, L~Gesa,
  F~Gibert, D~Giardini, C~Grimani, A~Grynagier, B~Guillaume, F~Guzmán,
  I~Harrison, M~Hewitson, D~Hollington, J~Hough, D~Hoyland, M~Hueller,
  J~Huesler, O~Jeannin, O~Jennrich, P~Jetzer, B~Johlander, C~Killow, X~Llamas,
  I~Lloro, A~Lobo, R~Maarschalkerweerd, S~Madden, D~Mance, I~Mateos, P~W
  McNamara, J~Mendes, E~Mitchell, D~Nicolini, D~Nicolodi, F~Pedersen,
  M~Perreur-Lloyd, A~Perreca, E~Plagnol, P~Prat, G~D Racca, B~Rais,
  J~Ramos-Castro, J~Reiche, J~A~Romera Perez, D~Robertson, H~Rozemeijer,
  J~Sanjuan, M~Schulte, D~Shaul, L~Stagnaro, S~Strandmoe, T~J Sumner, A~Taylor,
  D~Texier, C~Trenkel, D~Tombolato, S~Vitale, G~Wanner, H~Ward, S~Waschke,
  P~Wass, W~J Weber, and P~Zweifel.
\newblock The LISA Pathfinder interferometry—hardware and system testing.
\newblock \emph{Classical and Quantum Gravity}, 28\penalty0 (9):\penalty0
  094003, 2011.
\newblock URL \url{http://stacks.iop.org/0264-9381/28/i=9/a=094003}.

\bibitem[Dolesi et~al.(2003)Dolesi, Bortoluzzi, Bosetti, Carbone, Cavalleri,
  Cristofolini, DaLio, Fontana, Fontanari, Foulon, Hoyle, Hueller, Nappo,
  Sarra, Shaul, Sumner, Weber, and Vitale]{grs1}
R~Dolesi, D~Bortoluzzi, P~Bosetti, L~Carbone, A~Cavalleri, I~Cristofolini,
  M~DaLio, G~Fontana, V~Fontanari, B~Foulon, C~D Hoyle, M~Hueller, F~Nappo,
  P~Sarra, D~N~A Shaul, T~Sumner, W~J Weber, and S~Vitale.
\newblock Gravitational sensor for LISA and its technology demonstration
  mission.
\newblock \emph{Classical and Quantum Gravity}, 20\penalty0 (10):\penalty0 S99,
  2003.
\newblock URL \url{http://stacks.iop.org/0264-9381/20/i=10/a=312}.

\bibitem[Vitale et~al.(2002)]{ltpdefinitiondoc}
Stefano Vitale et~al.
\newblock {The LISA Technology Package on board SMART-2, The LTP Definition
  Document, Unitn-Int 10-2002/Rel. 1.3}, October 2002.
\newblock URL
  \url{http://www.rssd.esa.int/SP/LISAPATHFINDER/docs/Top_level_documents/LTP_DD.pdf}.

\bibitem[Wanner et~al.(2017)Wanner, Karnesis, and collaboration]{wanner2017}
Gudrun Wanner, Nikolaos Karnesis, and LISA~Pathfinder collaboration.
\newblock Preliminary results on the suppression of sensing cross-talk in LISA
  Pathfinder.
\newblock \emph{Journal of Physics: Conference Series}, 840\penalty0
  (1):\penalty0 012043, 2017.
\newblock URL \url{http://stacks.iop.org/1742-6596/840/i=1/a=012043}.

\bibitem[Congedo et~al.(2012)Congedo, Ferraioli, Hueller, De~Marchi, Vitale,
  Armano, Hewitson, and Nofrarias]{PhysRevD.85.122004}
G.~Congedo, L.~Ferraioli, M.~Hueller, F.~De~Marchi, S.~Vitale, M.~Armano,
  M.~Hewitson, and M.~Nofrarias.
\newblock Time domain maximum likelihood parameter estimation in LISA
  Pathfinder data analysis.
\newblock \emph{Phys. Rev. D}, 85:\penalty0 122004, Jun 2012.
\newblock \doi{10.1103/PhysRevD.85.122004}.
\newblock URL \url{https://link.aps.org/doi/10.1103/PhysRevD.85.122004}.

\bibitem[Nofrarias et~al.(2012)Nofrarias, Ferraioli, Congedo, Hueller, Armano,
  Diaz-Aguiló, Grynagier, Hewitson, and Vitale]{1742-6596-363-1-012053}
M~Nofrarias, L~Ferraioli, G~Congedo, M~Hueller, M~Armano, M~Diaz-Aguiló,
  A~Grynagier, M~Hewitson, and S~Vitale.
\newblock Parameter estimation in LISA Pathfinder operational exercises.
\newblock \emph{Journal of Physics: Conference Series}, 363\penalty0
  (1):\penalty0 012053, 2012.
\newblock URL \url{http://stacks.iop.org/1742-6596/363/i=1/a=012053}.

\bibitem[Karnesis et~al.(2014)Karnesis, Nofrarias, Sopuerta, Gibert, Armano,
  Audley, Congedo, Diepholz, Ferraioli, Hewitson, Hueller, Korsakova, McNamara,
  Plagnol, and Vitale]{PhysRevD.89.062001}
Nikolaos Karnesis, Miquel Nofrarias, Carlos~F. Sopuerta, Ferran Gibert, Michele
  Armano, Heather Audley, Giuseppe Congedo, Ingo Diepholz, Luigi Ferraioli,
  Martin Hewitson, Mauro Hueller, Natalia Korsakova, Paul~W. McNamara, Eric
  Plagnol, and Stefano Vitale.
\newblock Bayesian model selection for LISA Pathfinder.
\newblock \emph{Phys. Rev. D}, 89:\penalty0 062001, Mar 2014.
\newblock \doi{10.1103/PhysRevD.89.062001}.
\newblock URL \url{https://link.aps.org/doi/10.1103/PhysRevD.89.062001}.

\bibitem[et~al.()]{lambdamiscalibration}
M.~Armano et~al.
\newblock {in preparation}.

\bibitem[Vitale et~al.(2014)Vitale, Congedo, Dolesi, Ferroni, Hueller,
  Vetrugno, Weber, Audley, Danzmann, Diepholz, Hewitson, Korsakova, Ferraioli,
  Gibert, Karnesis, Nofrarias, Inchauspe, Plagnol, Jennrich, McNamara, Armano,
  Thorpe, and Wass]{stefanologarithmic}
Stefano Vitale, Giuseppe Congedo, Rita Dolesi, Valerio Ferroni, Mauro Hueller,
  Daniele Vetrugno, William~Joseph Weber, Heather Audley, Karsten Danzmann,
  Ingo Diepholz, Martin Hewitson, Natalia Korsakova, Luigi Ferraioli, Ferran
  Gibert, Nikolaos Karnesis, Miquel Nofrarias, Henri Inchauspe, Eric Plagnol,
  Oliver Jennrich, Paul~W. McNamara, Michele Armano, James~Ira Thorpe, and
  Peter Wass.
\newblock Data series subtraction with unknown and unmodeled background noise.
\newblock \emph{Phys. Rev. D}, 90:\penalty0 042003, Aug 2014.
\newblock \doi{10.1103/PhysRevD.90.042003}.
\newblock URL \url{http://link.aps.org/doi/10.1103/PhysRevD.90.042003}.

\bibitem[Armano et~al.(2016{\natexlab{b}})Armano, Audley, Auger, Baird,
  Binetruy, Born, Bortoluzzi, Brandt, Bursi, Caleno, Cavalleri, Cesarini,
  Cruise, Danzmann, de~Deus~Silva, Desiderio, Piersanti, Diepholz, Dolesi,
  Dunbar, Ferraioli, Ferroni, Fitzsimons, Flatscher, Freschi, Gallegos,
  Marirrodriga, Gerndt, Gesa, Gibert, Giardini, Giusteri, Grimani, Grzymisch,
  Harrison, Heinzel, Hewitson, Hollington, Hueller, Huesler, Inchauspé,
  Jennrich, Jetzer, Johlander, Karnesis, Kaune, Korsakova, Killow, Lloro, Liu,
  López-Zaragoza, Maarschalkerweerd, Madden, Mance, Martín, Martin-Polo,
  Martino, Martin-Porqueras, Mateos, McNamara, Mendes, Mendes, Moroni,
  Nofrarias, Paczkowski, Perreur-Lloyd, Petiteau, Pivato, Plagnol, Prat,
  Ragnit, Ramos-Castro, Reiche, Perez, Robertson, Rozemeijer, Rivas, Russano,
  Sarra, Schleicher, Slutsky, Sopuerta, Sumner, Texier, Thorpe, Tomlinson,
  Trenkel, Vetrugno, Vitale, Wanner, Ward, Warren, Wass, Wealthy, Weber,
  Wittchen, Zanoni, Ziegler, and Zweifel]{0264-9381-33-23-235015}
M~Armano, H~Audley, G~Auger, J~Baird, P~Binetruy, M~Born, D~Bortoluzzi,
  N~Brandt, A~Bursi, M~Caleno, A~Cavalleri, A~Cesarini, M~Cruise, K~Danzmann,
  M~de~Deus~Silva, D~Desiderio, E~Piersanti, I~Diepholz, R~Dolesi, N~Dunbar,
  L~Ferraioli, V~Ferroni, E~Fitzsimons, R~Flatscher, M~Freschi, J~Gallegos,
  C~García Marirrodriga, R~Gerndt, L~Gesa, F~Gibert, D~Giardini, R~Giusteri,
  C~Grimani, J~Grzymisch, I~Harrison, G~Heinzel, M~Hewitson, D~Hollington,
  M~Hueller, J~Huesler, H~Inchauspé, O~Jennrich, P~Jetzer, B~Johlander,
  N~Karnesis, B~Kaune, N~Korsakova, C~Killow, I~Lloro, L~Liu, J~P
  López-Zaragoza, R~Maarschalkerweerd, S~Madden, D~Mance, V~Martín,
  L~Martin-Polo, J~Martino, F~Martin-Porqueras, I~Mateos, P~W McNamara,
  J~Mendes, L~Mendes, A~Moroni, M~Nofrarias, S~Paczkowski, M~Perreur-Lloyd,
  A~Petiteau, P~Pivato, E~Plagnol, P~Prat, U~Ragnit, J~Ramos-Castro, J~Reiche,
  J~A~Romera Perez, D~Robertson, H~Rozemeijer, F~Rivas, G~Russano, P~Sarra,
  A~Schleicher, J~Slutsky, C~F Sopuerta, T~Sumner, D~Texier, J~I Thorpe,
  R~Tomlinson, C~Trenkel, D~Vetrugno, S~Vitale, G~Wanner, H~Ward, C~Warren, P~J
  Wass, D~Wealthy, W~J Weber, A~Wittchen, C~Zanoni, T~Ziegler, and P~Zweifel.
\newblock Constraints on LISA Pathfinder’s self-gravity: design requirements,
  estimates and testing procedures.
\newblock \emph{Classical and Quantum Gravity}, 33\penalty0 (23):\penalty0
  235015, 2016{\natexlab{b}}.
\newblock URL \url{http://stacks.iop.org/0264-9381/33/i=23/a=235015}.

\bibitem[Weber et~al.(2003)Weber, Bortoluzzi, Cavalleri, Carbone, Da~Lio,
  Dolesi, Fontana, Hoyle, Hueller, and Vitale]{SPIE}
William~J. Weber, Daniele Bortoluzzi, Antonella Cavalleri, Ludovico Carbone,
  Mauro Da~Lio, Rita Dolesi, Giorgio Fontana, C.~D. Hoyle, Mauro Hueller, and
  Stefano Vitale.
\newblock {Position sensors for flight testing of LISA drag-free control}.
\newblock \emph{Proc. SPIE Int. Soc. Opt. Eng.}, 4856:\penalty0 31--42, 2003.
\newblock \doi{10.1117/12.458564}.

\bibitem[Brandt and Fichter(2009)]{nico_lisa_symp}
N.~Brandt and W.~Fichter.
\newblock Revised electrostatic model of the LISA Pathfinder inertial sensor.
\newblock \emph{J. Phys.: Conf. Ser.}, 154:\penalty0 012008, 2009.

\bibitem[Cutler and Flanagan(1994)]{PhysRevD.49.2658}
Curt Cutler and \'Eanna~E. Flanagan.
\newblock Gravitational waves from merging compact binaries: How accurately can
  one extract the binary's parameters from the inspiral waveform?
\newblock \emph{Phys. Rev. D}, 49:\penalty0 2658--2697, Mar 1994.
\newblock \doi{10.1103/PhysRevD.49.2658}.
\newblock URL \url{https://link.aps.org/doi/10.1103/PhysRevD.49.2658}.

\bibitem[Nofrarias et~al.(2010)Nofrarias, R\"over, Hewitson, Monsky, Heinzel,
  Danzmann, Ferraioli, Hueller, and Vitale]{PhysRevD.82.122002}
M.~Nofrarias, C.~R\"over, M.~Hewitson, A.~Monsky, G.~Heinzel, K.~Danzmann,
  L.~Ferraioli, M.~Hueller, and S.~Vitale.
\newblock Bayesian parameter estimation in the second LISA Pathfinder mock data
  challenge.
\newblock \emph{Phys. Rev. D}, 82:\penalty0 122002, Dec 2010.
\newblock \doi{10.1103/PhysRevD.82.122002}.
\newblock URL \url{https://link.aps.org/doi/10.1103/PhysRevD.82.122002}.

\bibitem[R\"over(2011)]{rover1}
Christian R\"over.
\newblock Student-$t$ based filter for robust signal detection.
\newblock \emph{Phys. Rev. D}, 84:\penalty0 122004, Dec 2011.
\newblock \doi{10.1103/PhysRevD.84.122004}.
\newblock URL \url{http://link.aps.org/doi/10.1103/PhysRevD.84.122004}.

\bibitem[R\"{o}ver et~al.(2011)R\"{o}ver, Meyer, and Christensen]{rover2}
Christian R\"{o}ver, Renate Meyer, and Nelson Christensen.
\newblock Modelling coloured residual noise in gravitational-wave signal
  processing.
\newblock \emph{Classical and Quantum Gravity}, 28\penalty0 (1):\penalty0
  015010, 2011.
\newblock URL \url{http://stacks.iop.org/0264-9381/28/i=1/a=015010}.

\bibitem[Wanner(2010)]{Wanner2010thesis}
Gudrun Wanner.
\newblock \emph{Complex optical systems in space: numerical modelling of the
  heterodyne interferometry of LISA Pathfinder and LISA}.
\newblock PhD thesis, Leibniz University, Hannover, 2010.
\newblock URL
  \url{http://edok01.tib.uni-hannover.de/edoks/e01dh11/660137038.pdf}.

\bibitem[Tinto and Dhurandhar(2005)]{Tinto2005}
Massimo Tinto and Sanjeev~V. Dhurandhar.
\newblock Time-delay interferometry.
\newblock \emph{Living Reviews in Relativity}, 8\penalty0 (1):\penalty0 4, Jul
  2005.
\newblock ISSN 1433-8351.
\newblock \doi{10.12942/lrr-2005-4}.
\newblock URL \url{https://doi.org/10.12942/lrr-2005-4}.

\end{thebibliography}
\end{document}